\documentclass[twocolumn,trackchanges]{aastex701}

\begin{document}

\title{A Catalog of Giant Radio Source Candidates from TGSS ADR1 Using a Deep Learning--based Pipeline}

\author[orcid=0000-0002-3426-3269,sname='Lao']{Baoqiang Lao}
\affiliation{Guangxi Key Laboratory of Brain-Inspired Computing and Intelligent Chips, School of Electronic and Information Engineering, GuangXi Normal University, Guilin 541004, People's Republic of China}
\affiliation{Key Laboratory of Nonlinear Circuits and Optical Communications (GuangXi Normal University), Education Department of Guangxi Zhuang Autonomous Region, Guilin 541004, People's Republic of China}
\email[show]{lbq@gxnu.edu.cn}  

\author[sname='Liang']{Binghua Liang}
\affiliation{Guangxi Key Laboratory of Brain-Inspired Computing and Intelligent Chips, School of Electronic and Information Engineering, GuangXi Normal University, Guilin 541004, People's Republic of China}
\affiliation{Key Laboratory of Nonlinear Circuits and Optical Communications (GuangXi Normal University), Education Department of Guangxi Zhuang Autonomous Region, Guilin 541004, People's Republic of China}
\email{3244049149@qq.com} 

\author[orcid=0000-0003-0181-7656,sname='Guo']{Shaoguang Guo}
\affiliation{Shanghai Astronomical Observatory, Chinese Academy of Sciences, Shanghai 200030, People's Republic of China}
\affiliation{University of Chinese Academy of Sciences, Beijing 100049, People's Republic of China}
\affiliation{State Key Laboratory of Radio Astronomy and Technology, Chinese Academy of Sciences, Beijing 100101, People's Republic of China}
\email[show]{sgguo@shao.ac.cn}

\author[sname='Liu']{Junxiu Liu}
\affiliation{Guangxi Key Laboratory of Brain-Inspired Computing and Intelligent Chips, School of Electronic and Information Engineering, GuangXi Normal University, Guilin 541004, People's Republic of China}
\affiliation{Key Laboratory of Nonlinear Circuits and Optical Communications (GuangXi Normal University), Education Department of Guangxi Zhuang Autonomous Region, Guilin 541004, People's Republic of China}
\email{liujunxiu@gxnu.edu.cn} 

\author[orcid=0000-0002-4439-5580,sname='Yang']{Xiaolong Yang}
\affiliation{Shanghai Astronomical Observatory, Chinese Academy of Sciences, Shanghai 200030, People's Republic of China}
\email{yangxl@shao.ac.cn}

\author[orcid=0000-0002-1243-0476, sname='Zhao']{Rushuang Zhao}
\affiliation{School of Physics and Electronic Science, Guizhou Normal University, Guiyang 550001, People's Republic of China}
\email{201907007@gznu.edu.cn}

\author[sname='Kun']{Yang Kun}
\affiliation{Minzu Normal University of Xingyi, Xingyi 562400, People’s Republic of China}
\email{yangkun@xynun.edu.cn}


\begin{abstract}
We present a catalog of 5,595 giant radio source (GRS) candidates, defined by a largest (projected) linear size (LLS) exceeding 0.7 Mpc. Of these, 4,566 would be new discoveries if confirmed. These candidates were identified through a systematic and automated search of the first Alternative Data Release of the TIFR GMRT Sky Survey (TGSS ADR1) using a deep learning--based pipeline that incorporates a radio sources detection model and the likelihood ratio method. Across the full sample, the LLS reaches up to $\sim$4.5~Mpc, with redshifts ranging from $z=0.056$ to at least $z=2.385$. The catalog comprises 4,210 giant radio galaxies and 1,030 giant radio quasars, with the remaining 355 candidates currently unclassified. We systematically analyze the distributions of their key physical properties, including spectral index, bending angle (BA), radio power ($P_{150}$), and $r$-band absolute magnitude. Our results show that GRS candidates are predominantly straight (a median BA of $11.0^\circ$) and follow an ``L-shaped" BA--LLS distribution. The radio power vs. LLS ($P_{150}$--$D$) diagram reveals a concentration at $P_{150} < 10^{28}$ W Hz$^{-1}$, while rare high-power outliers ($>10^{28.5}$ W Hz$^{-1}$) with LLS $>$ 1 Mpc challenge self-similar evolutionary models. Additionally, 286 GRS candidates are found to be associated with cataloged galaxy clusters.

\end{abstract}



\section{Introduction}

Active galactic nuclei (AGNs), powered by accretion onto supermassive black holes, act as cosmic engines that convert gravitational energy into intense radiation and powerful outflows. Radio-loud AGNs launch relativistic jets that inflate enormous radio lobes, forming radio sources. The giant members of this class are of special interest. Giant radio sources (GRSs), encompassing both giant radio galaxies (GRGs) and giant radio quasars (GRQs), are conventionally defined as those with a largest (projected) linear size (LLS) exceeding 0.7 Mpc ($\sim$2.3 million light-years) \citep{1974Natur.250..625W, 1999MNRAS.309..100I}. They rank among the largest single astrophysical structures known in the universe, with extents far surpassing their host galaxies and even comparable to the scales of galaxy clusters \citep[e.g.,][]{2001A&A...371..445M}, with the largest known example spanning approximately 7 Mpc \citep{2024Natur.633..537O}.
GRGs are typically hosted by elliptical galaxies, whereas GRQs possess bright quasar cores; together, they provide a natural cosmological laboratory for studying long-term AGN feedback and extreme physical environments.

The ultimate challenge in understanding AGN feedback lies in explaining how their jets maintain collimation and propagate across megaparsec scales through the tenuous intergalactic or intracluster medium, injecting vast amounts of energy that critically shape galaxy evolution and cluster thermodynamics \citep[e.g.,][]{2021A&A...647A...3B,2020A&A...635A...5D,2025A&A...697A.147D}. A deeper understanding of this physical process is intrinsically linked to the morphological classification of radio sources. Radio sources are primarily categorized as either compact sources (CS) or extended sources. Extended sources are further classified morphologically into: Fanaroff--Riley Type I/II (FR I/II) sources, core-jet (CJ) sources, bent-tail (BT) sources, as well as X-shaped and S-shaped sources \citep{2011ApJS..194...31P}. The latter categories typically involve jets that are distorted by dynamical interactions with the surrounding environment. Among them, FR I sources are typically edge-darkened, a morphology attributed to a lack of a ``second'' particle acceleration event as occurs in FR II hotspots, and to their jets' relativistic electrons becoming ``dilute'' or spiraling in weak magnetic fields. In contrast, FR II sources typically exhibit an edge-brightened morphology, with highly collimated jets terminating in bright hotspots, where the synchrotron energy loss timescale is inversely proportional to the Lorentz factor; thus, the highest-energy electrons radiate intensely and produce the characteristic edge-brightened morphology \citep{1974MNRAS.167P..31F}.
Notably, GRSs predominantly belong to the FR II class \citep[e.g.,][]{2024RAA....24c5021L}. Investigating the propagation mechanisms and energy deposition efficiency of these different jet types is key to unveiling the diversity and extremes of AGN feedback.

The systematic search for GRSs has evolved, but it still relies heavily on manual identification. The first generation of large-scale surveys, such as the NRAO VLA Sky Survey (NVSS) and the Faint Images of the Radio Sky at Twenty-cm (FIRST) survey, facilitated the creation of early catalogs \citep[e.g.,][]{2016ApJS..224...18P,2018ApJS..238....9K,2020MNRAS.499...68T}. However, limited by their surface brightness sensitivity, these surveys missed a significant number of sources with diffuse, low-brightness lobes, resulting in samples that are severely incomplete and subject to strong observational selection biases. The new generation of low-frequency, wide-field surveys, such as the LOFAR Two-metre Sky Survey \citep[LoTSS;][]{2017A&A...598A.104S} and the Rapid ASKAP Continuum Survey \citep[RACS;][]{2020PASA...37...48M}, with their superior sensitivity, has enabled an exponential growth in the number of known GRGs and GRQs. This is evidenced by discoveries such as large numbers of new sources in specific sky regions \citep{2020A&A...635A...5D,2020A&A...642A.153D,2022MNRAS.515.2032S,2023A&A...672A.163O,2024A&A...686A..21S,2024A&A...691A.185M} and the reporting of 178 GRSs from RACS low-frequency data \citep{2021Galax...9...99A}. Notably, \cite{2025A&A...699A.257A} presented a comprehensive sample of 142 GRGs exceeding 3 Mpc (including 69 new identifications) through a systematic search of modern radio surveys to investigate their physical properties and cluster environments, finding that these extreme sources are largely indistinguishable from their smaller counterparts. The GRS samples in the aforementioned studies were mostly identified through visual inspection. While automated methods based on supervised \citep[e.g.,][]{2022MNRAS.510.4504T} and unsupervised \citep[e.g.,][]{2021A&A...645A..89M} learning have been proposed, the efficacy of such techniques in GRS searches had not yet been rigorously quantified until the recent work of \cite{2024A&A...691A.185M}. Their machine learning--driven pipeline successfully identified and confirmed thousands of new GRSs, demonstrating a significant step forward in automating large-scale GRS searches.

In this context, the Tata Institute of Fundamental Research (TIFR) Giant Metrewave Radio Telescope (GMRT) Sky Survey (TGSS) \citep{2014A&A...562A.108S,2017A&A...598A..78I}, which covers approximately 90\% of the northern sky, is a valuable resource. The low observing frequency of this survey makes it particularly sensitive to the steep-spectrum lobes of GRSs, whose spectra are steeper than those of smaller radio sources, and these lobes are thus brighter at low frequencies \citep[e.g.,][]{2022MNRAS.515.2032S,2024ApJS..273...30B}. The larger sky coverage of TGSS, combined with an observing frequency closely matched to that of LoTSS (150~MHz vs. 144~MHz), enables direct comparison and complementarity with existing low-frequency GRS samples dominated by LoTSS. The potential of TGSS has been recently demonstrated \citep{2024ApJS..273...30B,2025ApJS..281...34M}. However, current search methodologies still face two fundamental bottlenecks: low search efficiency coupled with constraints on sample size, and the challenge of precisely identifying host galaxies. First, when confronted with large-scale survey data, manual methods relying on visual inspection suffer from slowness, a lack of scalability, and subjectivity, 
and thus lead to reduced reproducibility and a smaller final sample size, even though such methods may be better at identifying sources that are several tens of arcminutes long, overlap with many other sources, or are easily confused with diffuse emission like galaxy clusters mimicking source lobes. Second, for GRSs, the faint radio core is often offset from the extended lobe structures by arcminute-scale distances. This large angular separation, which increases with the physical size of the jet system, introduces numerous intervening galaxies that can be mistakenly identified as the host, thereby complicating the correct association of the radio emission with an optical/infrared host galaxy (for GRGs) or quasar (for GRQs). Such misidentifications can lead to substantial errors in estimating key physical parameters, such as redshift and LLS \citep{2020ApJS..247...53K}. Therefore, developing a fully automated, high-precision pipeline to discover GRS is crucial. Such a pipeline must be capable of processing large volumes of survey data to fully exploit the scientific potential of the TGSS and future Square Kilometre Array \citep[SKA;][]{2009IEEEP..97.1482D} era surveys.

This paper proposes an automatic method for identifying GRSs from TGSS data using a deep learning--based pipeline and performs a statistical analysis of their physical properties. This paper is organized as follows. Section~\ref{sec:data_and_method} describes the data and methods. The results are presented in Section~\ref{Sec:Results}. Finally, we provide a discussion and summary in Sections~\ref{Sec:Disc} and \ref{Sec:Conc}. Throughout this paper, we adopt a standard flat $\Lambda$CDM cosmological model with $H_0 = 70 \,{\rm km}\,{\rm s}^{-1}\,{\rm Mpc}^{-1}$ and $\Omega_{m,0} = 0.3$. For the radio spectral index, we follow the convention $F_\nu \propto \nu^\alpha$, where $F_\nu$ is the flux density at frequency $\nu$.

\section{Data and GRS Search Method} \label{sec:data_and_method}
\subsection{The TGSS ADR1 Data}
TGSS is a radio survey that produced a large-scale map of the sky at 150 MHz using the GMRT in India. In this paper, we use images and catalog data from the first Alternative Data Release of the TGSS \citep[TGSS ADR1;][]{2017A&A...598A..78I}, observed between April 2010 and March 2012. 
TGSS ADR1 includes continuum stokes I images covering 99.5\% of the radio sky north of -53$^\circ$ declination (36,900 deg$^2$, or 90\% of the full sky). The survey achieves an angular resolution of 25$''$$\times$25$''$ north of 19$^\circ$ declination and 25$''$$\times$$25''$/cos($\delta$-19$^\circ$) south of 19$^\circ$, with a median noise level of 3.5 mJy/beam. The data release comprises 5,336 mosaic images, each spanning 5$^\circ$$\times$5$^\circ$ with a resolution of 6.2$''$ per pixel and covering the entire survey area. The TGSS ADR1 radio source catalog provides positions, flux densities, sizes, and 623,604 sources detected at a 7 sigma significance level. The cataloged sources represent individual components of radio galaxies, quasars (i.e., quasi-stellar objects, QSOs), or star-forming galaxies, though no cross-component association or morphological classification has been applied. The large low-frequency sky coverage of TGSS ADR1 has facilitated the search for and study of diverse morphological classes of radio galaxies \citep[e.g.,][]{2022MNRAS.516..372B,2022MNRAS.512.4308B}.

\begin{figure*}[ht!]
\centering
\includegraphics[scale=0.85]{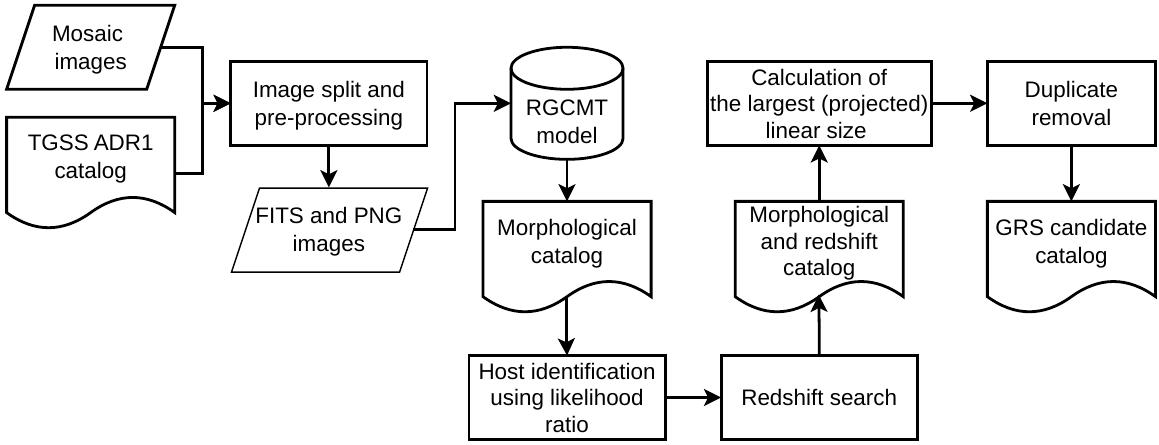}
\caption{Architecture of the proposed deep learning--based pipeline for automated detection of giant radio sources (GRSs).}.
\label{fig:auto_grg_det_pipe}
\end{figure*}

\subsection{Automated Detection of GRSs}
In this paper, we propose a deep learning--based pipeline to automate the detection of GRSs in the TGSS ADR1 images. The pipeline architecture, depicted in Figure~\ref{fig:auto_grg_det_pipe}, consists of six main steps: (1) image split and preprocessing, (2) localization and classification of radio sources, (3) infrared counterpart (host) identification, (4) redshift search, (5) calculation of the LLS, and (6) duplicate removal.

\begin{figure}[ht!]
\centering
\includegraphics[scale=0.3]{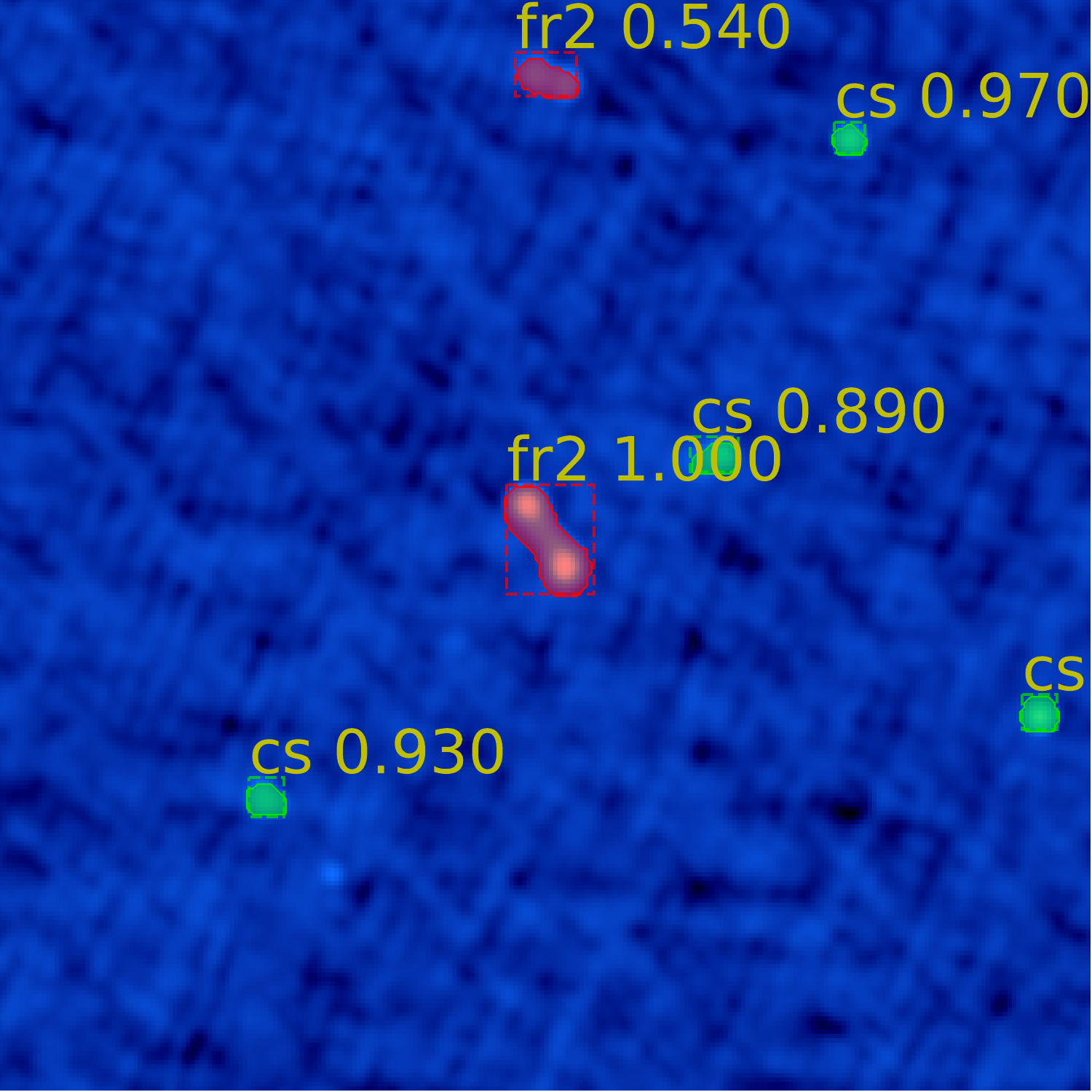}
\caption{RGCMT model detection results for a split TGSS image. Two FR II (red) and four CS (green) sources are identified, along with their labels (fr2: FR II, cs: CS), bounding boxes, and segmentation masks. The corresponding confidence scores (ranging from 0.0 to 1.0) are labeled for each detected source.
\label{fig:RGCMT_detect_example}}
\end{figure}

GRSs are predominantly observed to exhibit FR II morphology, while less than 10\% show FR I features. This study employs a deep learning model for radio source localization and classification: the Radio Galaxy Classification with Mask Transfiner \citep[RGCMT;][]{2023A&C....4400728L}, which can classify CS, FR I, FR II, CJ, and BT radio sources, while providing their bounding box coordinates, pixel-wise masks, confidence score, class label, centroid coordinates, and flux densities. From the TGSS ADR1 mosaics, our pipeline (based on this model) initially identifies sources of these various morphological types, from which we select only those classified as FR I or FR II for detailed examination. Although BT sources can in rare cases also be GRSs, we exclude them from this analysis because their complex morphology leads to lower host galaxy identification accuracy (via our automated method below) compared to the structurally simpler, slightly bent FR I and FR II sources, for which the host galaxy and geometric centroid are closely coincident. Since the original TGSS ADR1 mosaic images have large angular extents (5$^\circ$$\times$5$^\circ$), the RGCMT model exhibits high misclassification rates when processing full-field images directly. Therefore, we split the mosaics into $250\times250$-pixel FITS images (about 25.83$'\times$25.83$'$), centered on TGSS ADR1 catalog coordinates, and then convert them to logarithmic-scaled RGB PNGs using SAOImageDS9 \citep{2003ASPC..295..489J} with the `cool' colormap. This colormap matches the RGCMT training set, ensuring model transferability to TGSS ADR1. Following the splitting of mosaic images and subsequent preprocessing, a total of 623,604 FITS and PNG format images were generated. These were then processed by the RGCMT model to detect radio sources and produce a morphological catalog. Figure~\ref{fig:RGCMT_detect_example} shows the RGCMT model detection results for one of the split images. In this example, two FR II and four CS sources are identified and labeled with bounding boxes and masks, accompanied by confidence scores ranging from 0.0 to 1.0. For all split images, the RGCMT model detected 3,306,358 CS, 54,037 FR I, 170,352 FR II, 20,693 BT, and 215,500 CJ sources. It should be noted that the detections contain duplicates.

The third step of our pipeline focuses on identifying infrared counterparts for FR I and FR II sources within the morphological catalog. Given the substantial sample size, traditional host identification through visual inspection of radio--optical overlays \citep[e.g.,][]{2025ApJS..276...46L,2026ApJS..283...11L} becomes impractical. We instead implement an automated approach using the likelihood ratio (LR) method \citep[e.g.,][]{1992MNRAS.259..413S,2012MNRAS.423..132M}, which statistically determines the most probable optical/infrared counterpart for each radio source by evaluating positional uncertainties, source densities, and brightness distributions. This methodology has been successfully deployed in major radio surveys including the LOw Frequency ARray (LOFAR) Two-metre Sky Survey (LoTSS) \citep{2019A&A...622A...2W,2023A&A...678A.151H}, the MeerKAT International GHz Tiered Extragalactic Exploration (MIGHTEE) survey \citep{2024MNRAS.527.3231W}, and the Very Large Array Sky Survey (VLASS) \citep{2023ApJS..267...37G}. Our implementation follows the optimized framework described in \citet{2023ApJS..267...37G}, selected for its demonstrated reliability in handling large-scale catalogs.

We searched for infrared counterparts associated with our FR I/II sources using mid-infrared data from the Wide-field Infrared Survey Explorer \citep[WISE;][]{2010AJ....140.1868W}, taking advantage of its all-sky coverage in the 3.4 $\mu$m ($W1$) band. The AllWISE catalog \citep{2012yCat.2311....0C, 2013wise.rept....1C} provides astrometric precision better than 0$\farcs$5 even for faint sources, with a point spread function of 6$\farcs$1 in $W1$. The LR method proves particularly valuable for our FR I and FR II sources given the relatively low resolution of the TGSS data, as it effectively handles cases where poor-resolution radio observations may have multiple potential infrared counterparts \citep{2012MNRAS.423..132M}.

Following the method in \citet{2023ApJS..267...37G}, we determine the LR for all possible matches using the $W1$-band magnitude information for the AllWISE sources, defined as \citep{2012MNRAS.423..132M}:
\begin{equation}
LR = \frac{q(W1)f(d)}{n(W1)},
\label{eq:lr}
\end{equation}
where, $q(W1)$ signifies the probability density of a radio source being associated with an AllWISE counterpart at a specific $W1$-band magnitude; $f(d)$ denotes the probability distribution function for the radial separation between the radio source at its centroid coordinate and its AllWISE counterpart; and $n(W1)$ represents the sky distribution of background AllWISE sources as a function of $W1$-band magnitude. In this work, we first query the AllWISE catalog using a search radius of 125$''$ (5 times the angular resolution of 25$''$) to compile an initial set of potential counterparts for our FR I/FR II sources at its centroid coordinate. We then apply a signal-to-noise (S/N) cut of S/N$\geq$5 in at least one band to filter the AllWISE matches before performing the LR calculation. Equation \ref{eq:lr} quantifies the likelihood of a physical association between the radio cores of FR I/II sources and their infrared counterparts. This statistical measure is employed to identify the most probable IR counterparts of FR I/II hosts, thereby reducing contamination from spurious matches. We therefore characterize the reliability of each counterpart association as follows:
\begin{equation}
R_i = \frac{{L{R_i}}}{{\sum\nolimits_{j = 1}^N {L{R_j} - (1 - {Q_0})} }},
\label{eq:reliability}
\end{equation}
where $N$ represents the number of possible counterparts for a given radio source in the AllWISE catalog, and the $i$-th counterpart is denoted as AllWISE$_i$. $Q_0$ is the estimated fraction of radio sources with a true AllWISE counterpart \citep{2012MNRAS.423.2407F}. For any FR I/II source, the sum of reliabilities satisfies $\sum_{i=1}^N R_i = 1$. In this work, we identify the host for each FR I/II source by selecting the AllWISE match with the highest reliability $R_i$.

After identifying the infrared counterparts, we proceeded to obtain their redshifts. Using the host positions from AllWISE, we first retrieved spectroscopic redshifts by cross-matching with the following catalogs within a 1$''$ radius: the first data release of the Dark Energy Spectroscopic Instrument \citep[DESI DR1;][]{2026AJ....171..285D}, the seventeenth data release of the Sloan Digital Sky Survey \citep[SDSS DR17;][]{2022ApJS..259...35A}, the fourth data release of the Galaxy And Mass Assembly survey \citep[GAMA DR4;][]{2022MNRAS.513..439D}, the 2dF Galaxy Redshift Survey \citep[2dFGRS;][]{2001MNRAS.328.1039C}, the 6dF Galaxy Survey \citep[6dFGS;][]{2009MNRAS.399..683J}, the WiggleZ Dark Energy Survey \citep{2018MNRAS.474.4151D}, the Two Micron All Sky Survey Redshift Survey \citep[2MRS;][]{2012ApJS..199...26H}, and the NASA/IPAC Extragalactic Database (NED)\footnote{\url{https://ned.ipac.caltech.edu/}}. For those AllWISE hosts for which no spectroscopic redshift was found in the above catalogs, we then obtained a photometric redshift from the DESI Legacy Imaging Surveys Data Release 10 (DESI LS DR10). Finally, we incorporated the obtained redshifts into the morphological catalog to construct the combined morphological and redshift catalog.

\begin{figure}[ht!]
\centering
\includegraphics[scale=0.5]{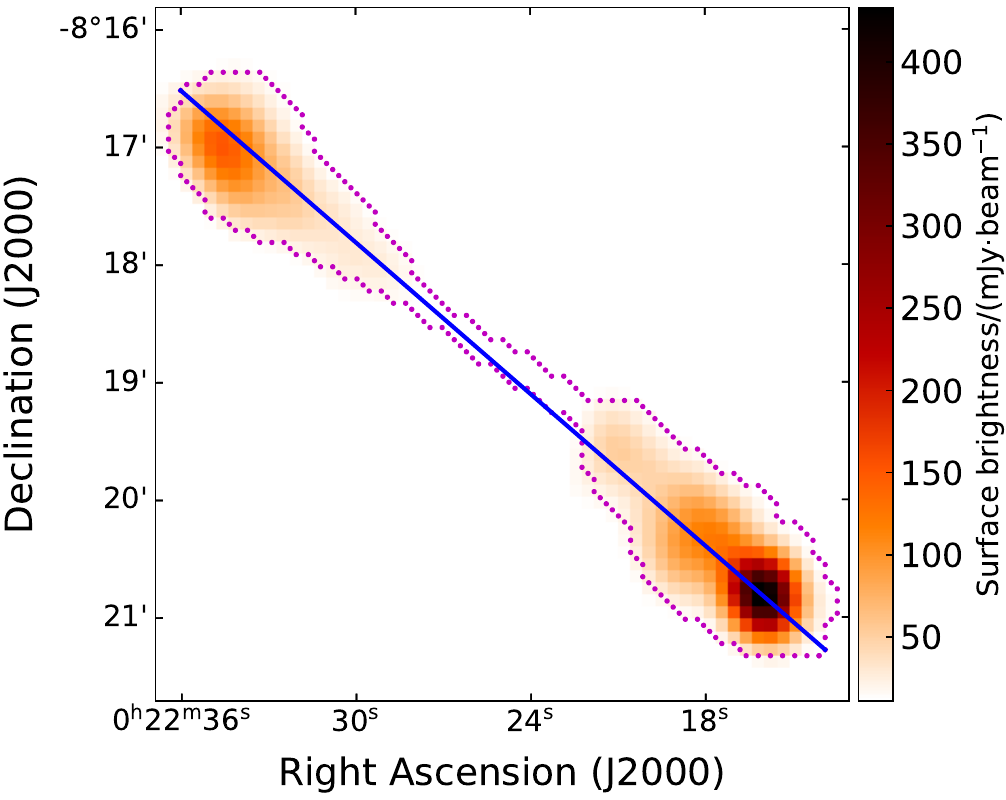}
\caption{Example of largest angular size (LAS) measurement for an FR II source (J0022-0818), determined by the longest separation between vertices in the purple polygon derived from the source's predicted mask. The blue line segment's length is the LAS.  
\label{fig:las_example}}
\end{figure}

In the next step, to identify GRSs, we calculated the LLS of the FR I and FR II sources.
This process began with an automated measurement of their largest angular sizes (LAS), as illustrated in Figure \ref{fig:las_example}. First, polygonal vertex coordinates (purple points in the figure) were derived from each source's predicted mask. These masks were generated by applying a 3$\sigma$ threshold (where $\sigma$ is the local RMS noise) to the image cutout enclosing the source. The LAS was then taken as the straight-line distance between the two farthest vertices of this polygon, shown by the blue line in the figure. Finally, the LLS of the sources was calculated based on their LAS measurements and redshifts, and sources with LLS less than 0.7 Mpc were removed. As mentioned earlier, the radio sources detected by the RGCMT model contain numerous duplicates due to overlapping regions in the split images. In the final step (i.e., the duplicate removal step), we performed an internal cross-match of the FR I and FR II sources that satisfied the LLS threshold, adopting a matching radius of 25$''$. For each matched group, we retained only the row with the highest confidence score that also exceeded 0.5.
This procedure yielded the final GRS candidate catalog containing 5,595 sources.

\section{Results} \label{Sec:Results}
\subsection{Our GRS Candidate Catalog: Source Properties and Statistics} \label{sec:catalog}
\begin{figure*}
    \centering
	\includegraphics[scale=0.25]{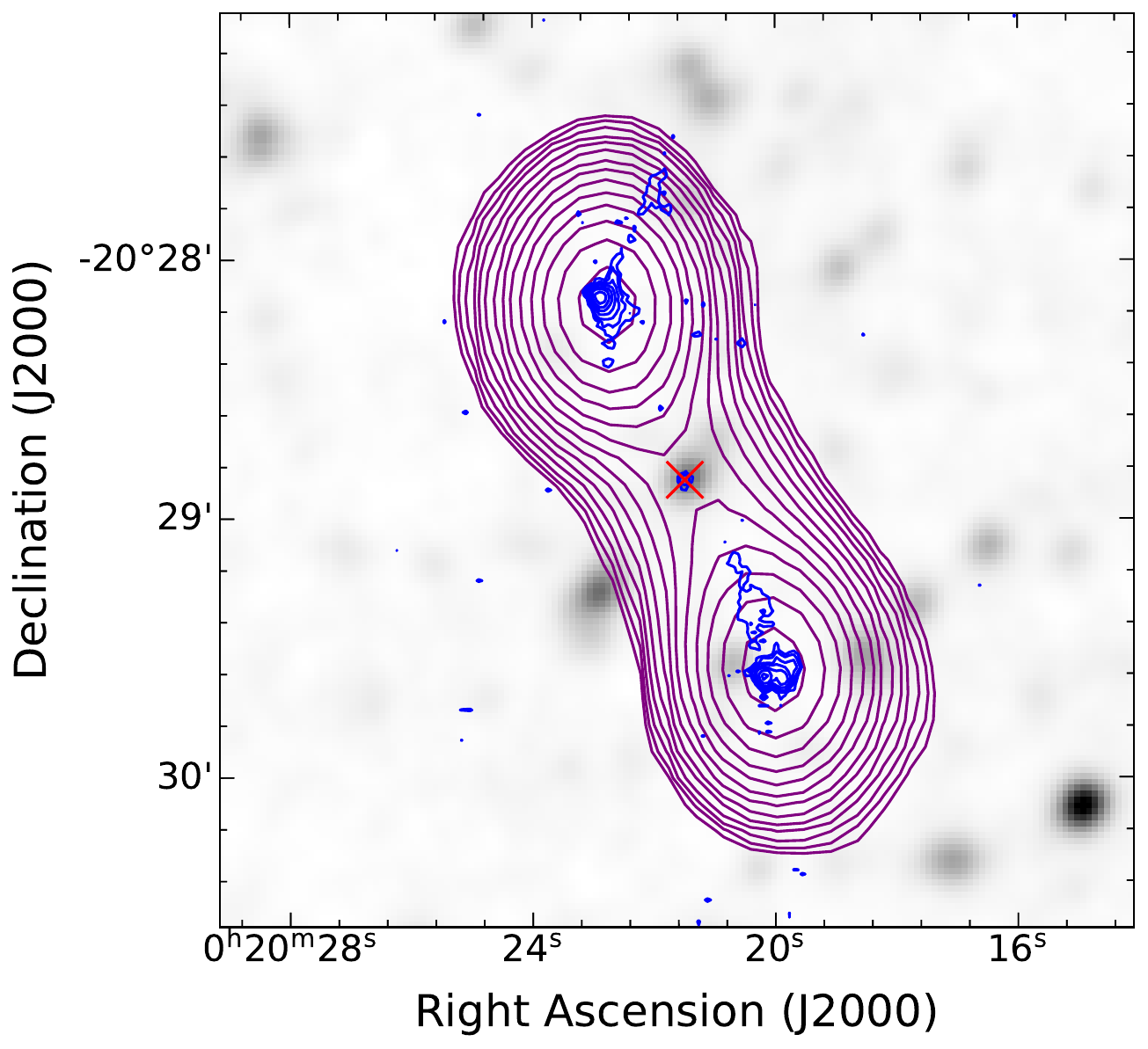} 
    \hspace{6mm}
    \includegraphics[scale=0.25]{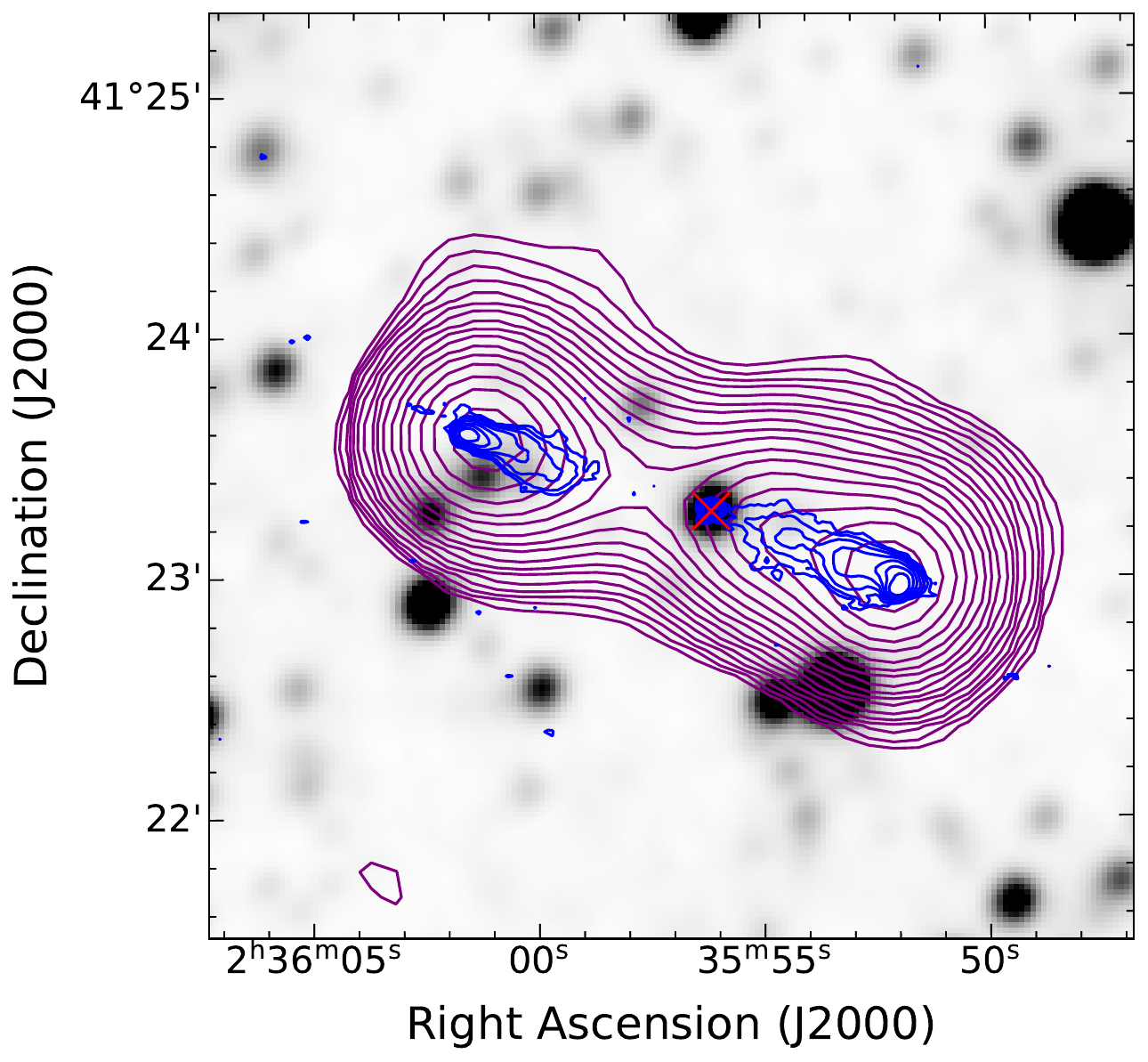} 
    \hspace{6mm}
    \includegraphics[scale=0.25]{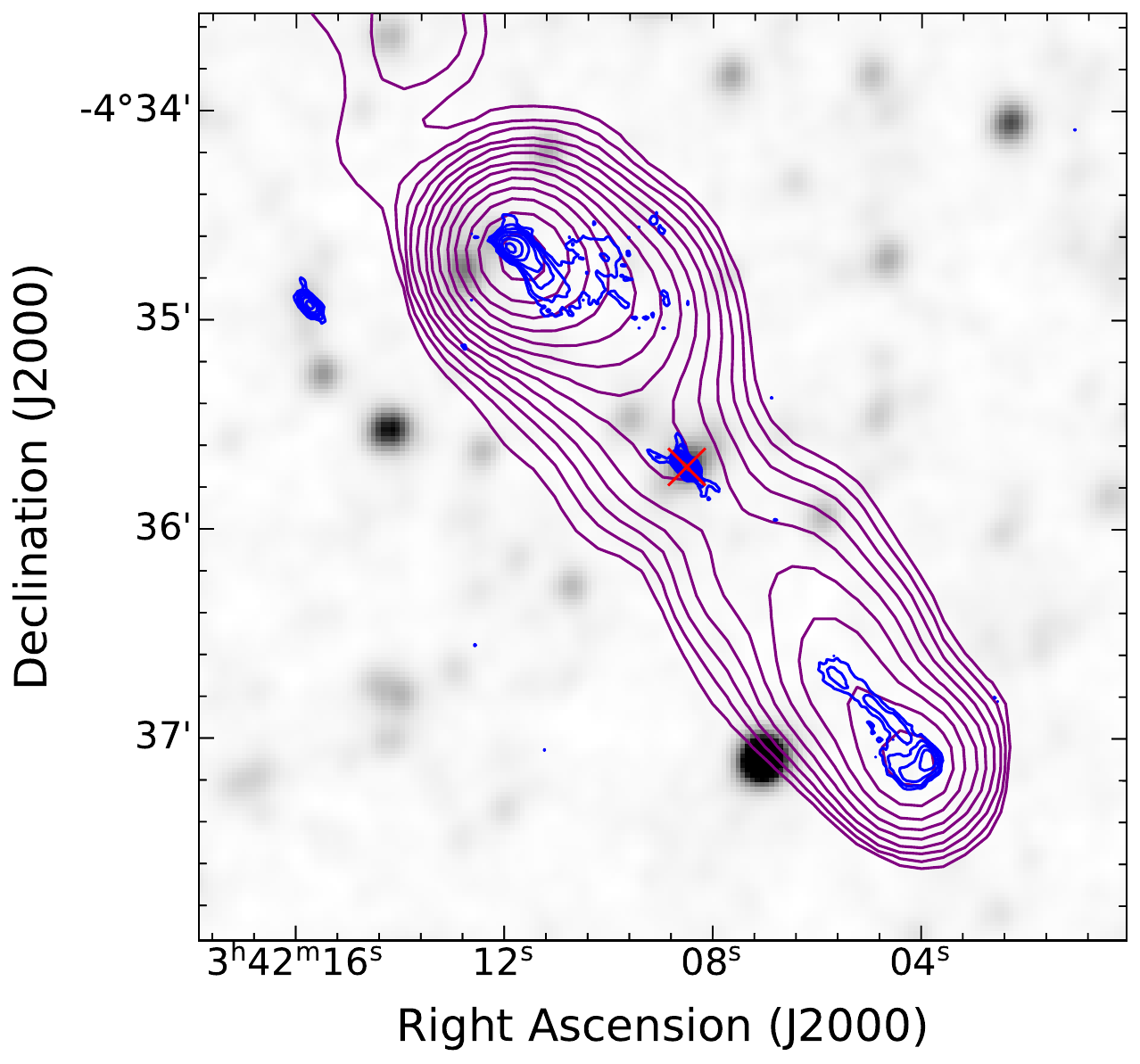} 
    \hspace{6mm}
    \includegraphics[scale=0.25]{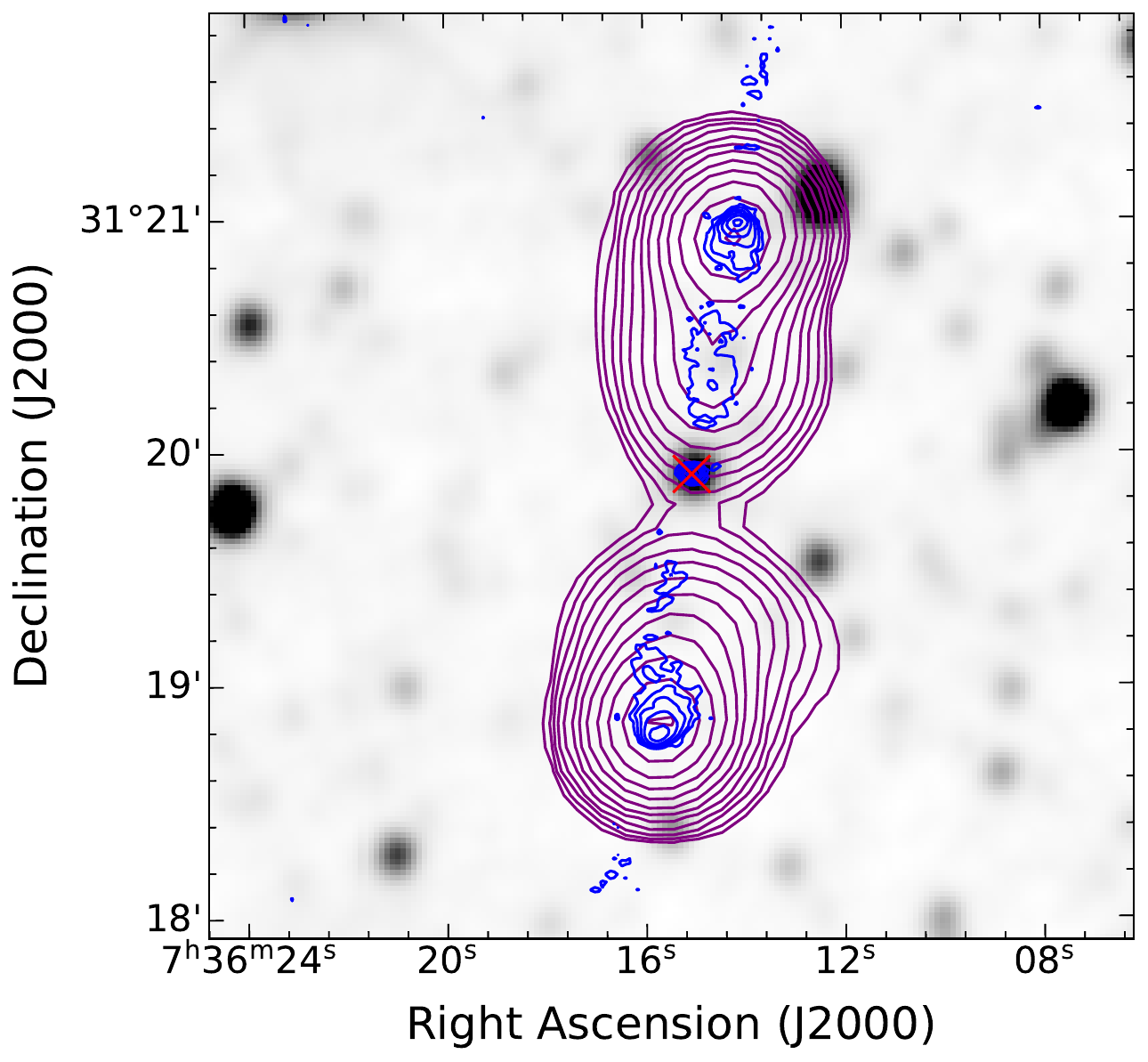} 
    \hspace{4mm}
    \includegraphics[scale=0.25]{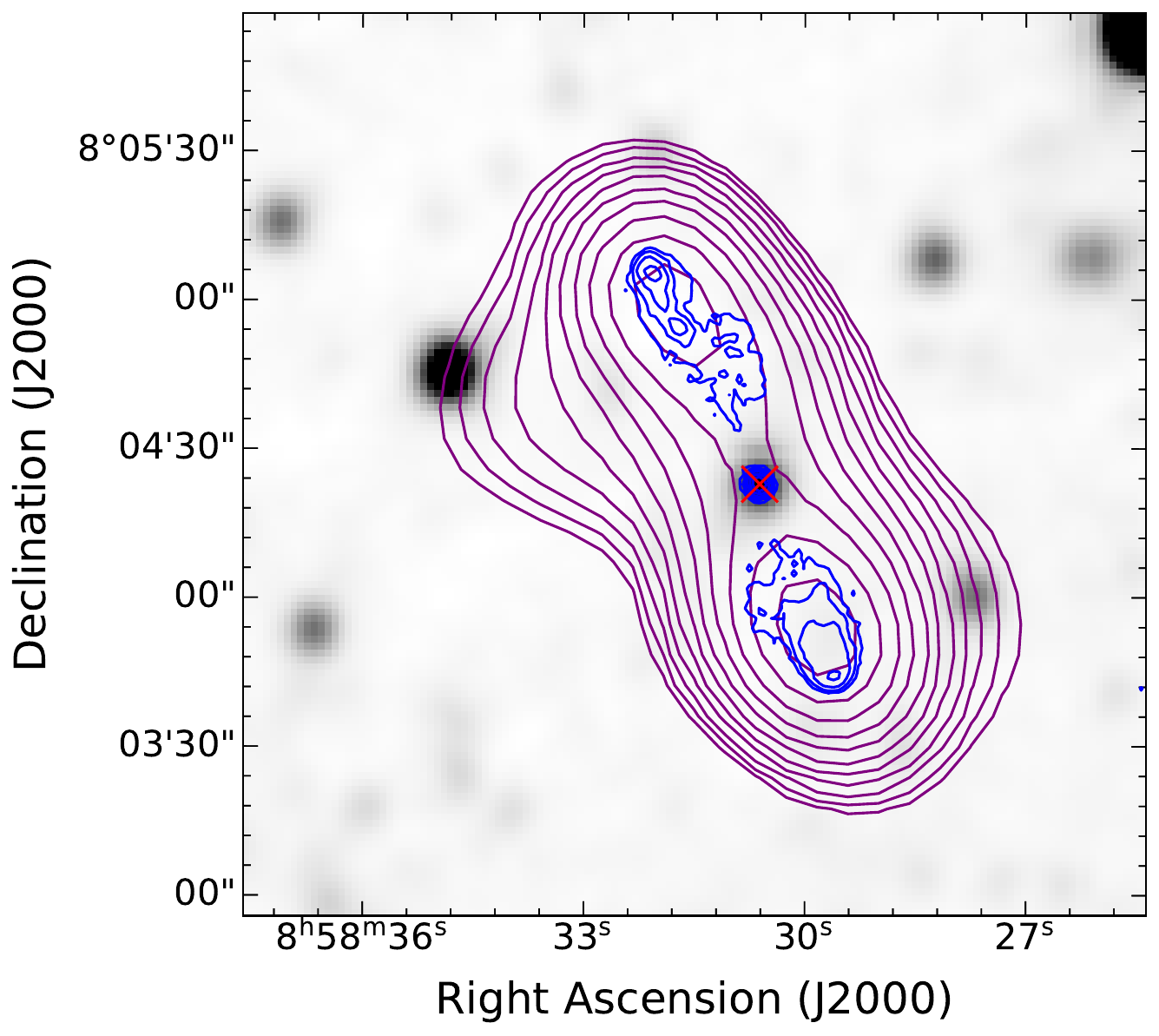} 
    \hspace{4mm}
    \includegraphics[scale=0.25]{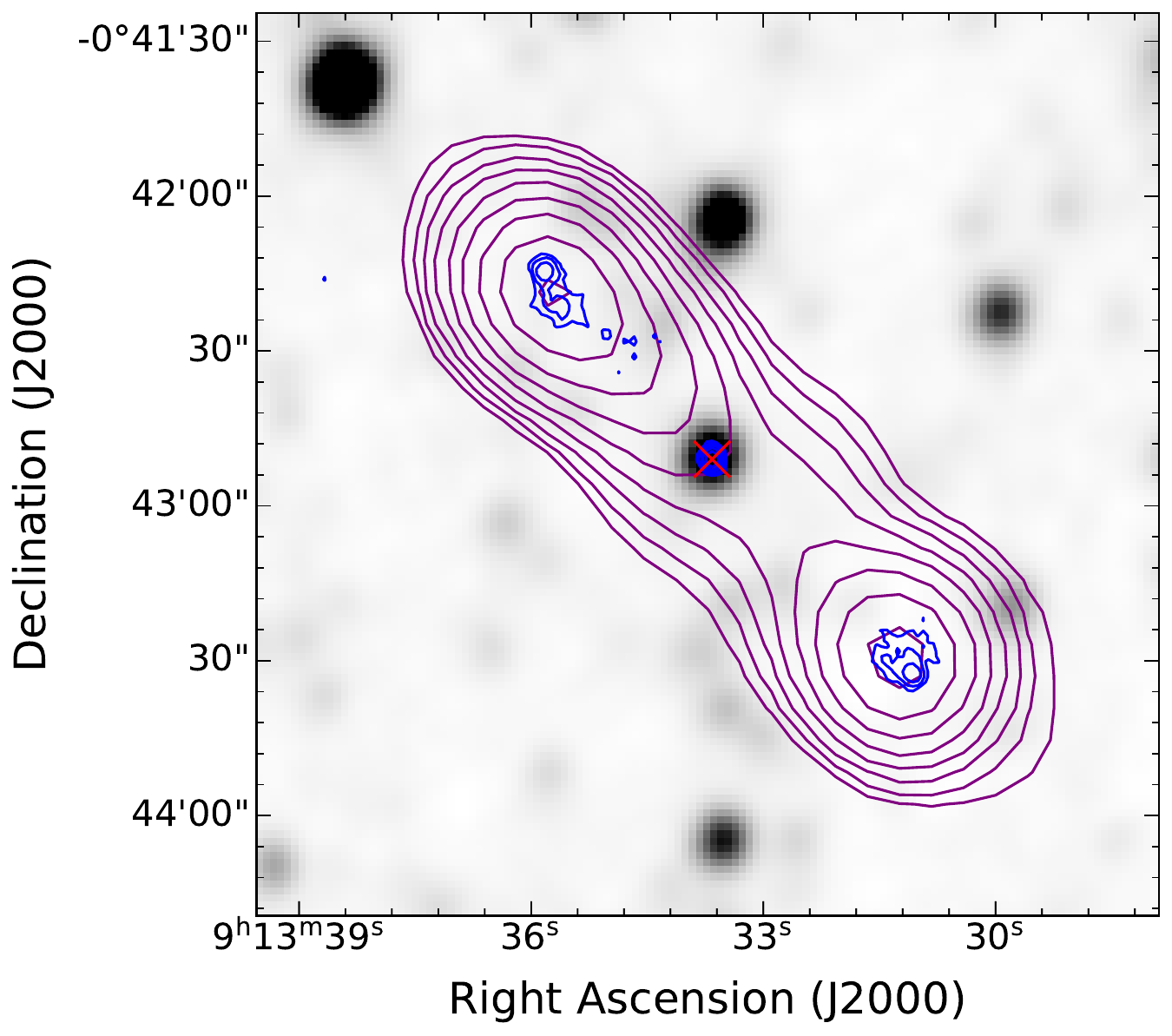} 
    \hspace{14mm}
    \includegraphics[scale=0.25]{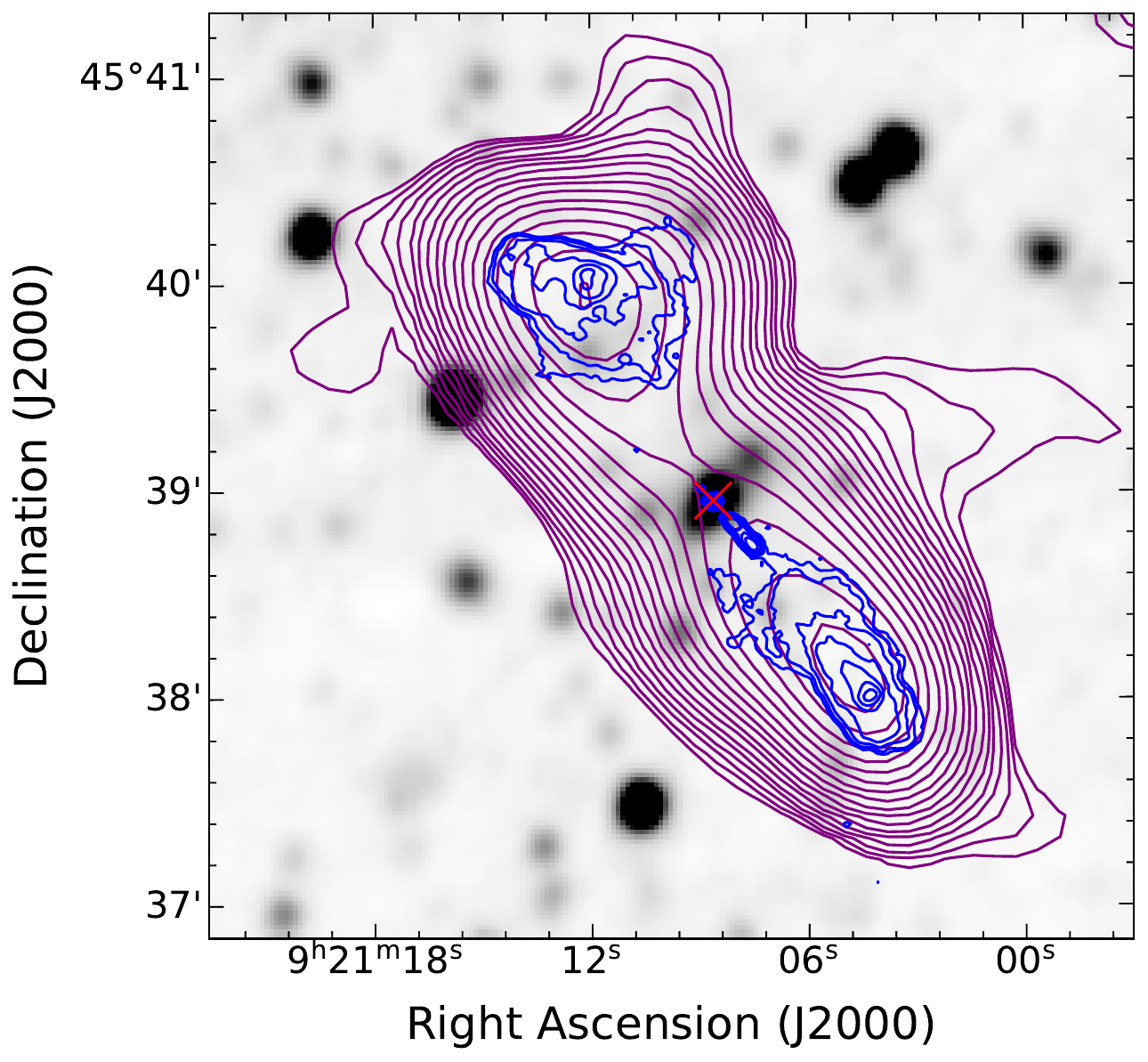}
    \hspace{6mm}
    \includegraphics[scale=0.25]{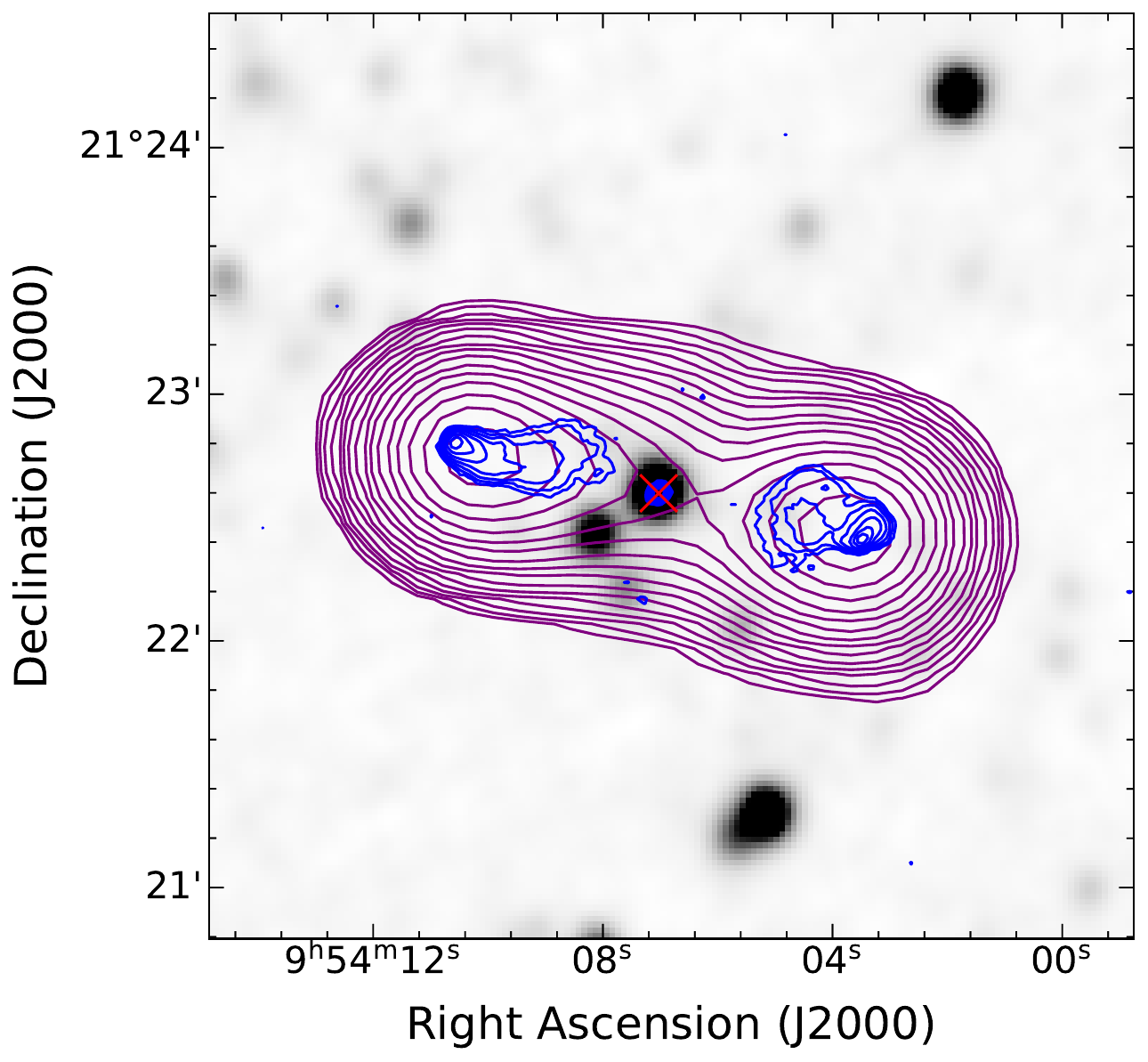} 
    \hspace{6mm}
    \includegraphics[scale=0.25]{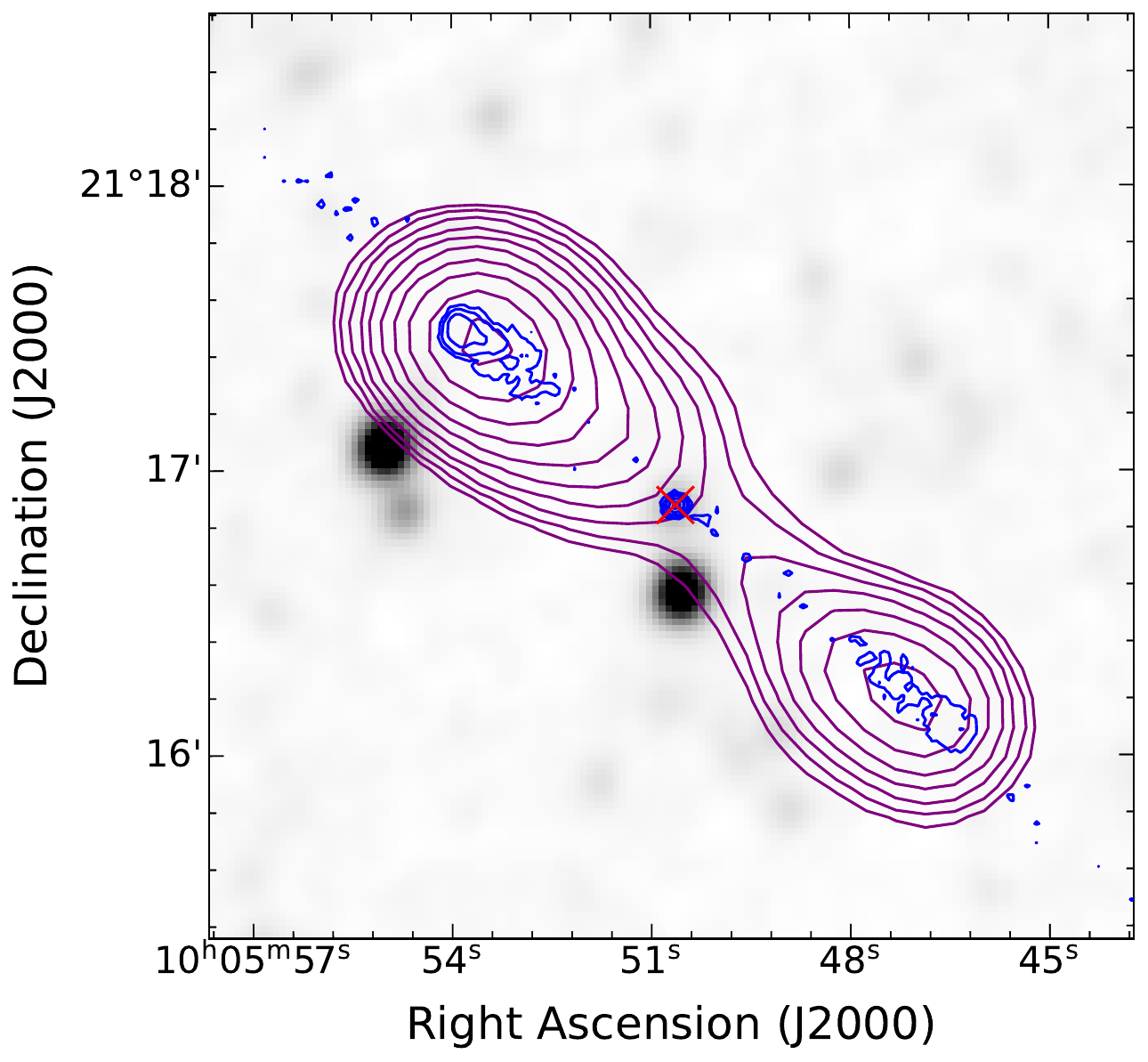} 
    \hspace{10mm}
    \includegraphics[scale=0.25]{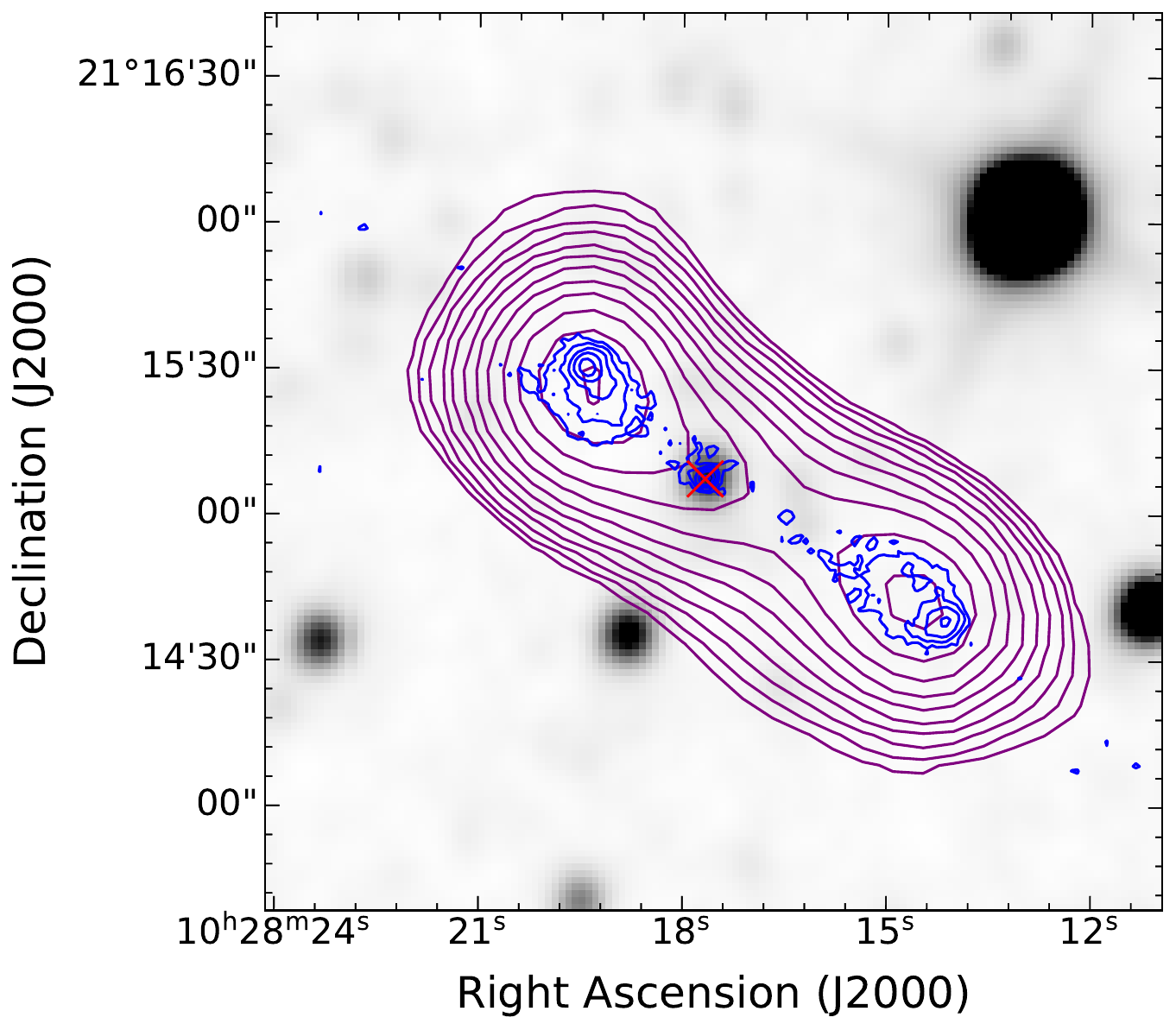} 
    \hspace{3mm}
    \includegraphics[scale=0.25]{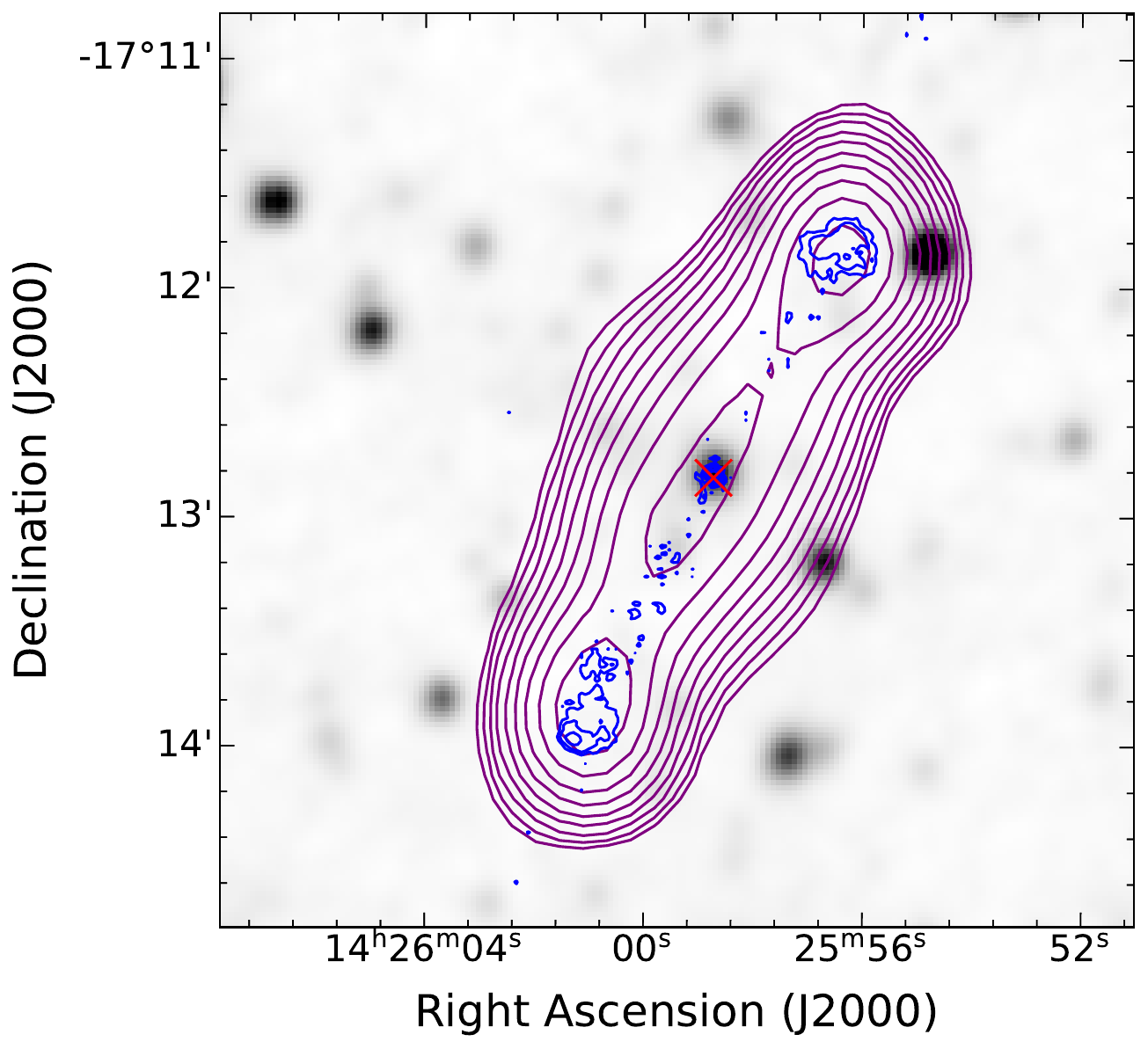}
    \hspace{3mm}
    \includegraphics[scale=0.25]{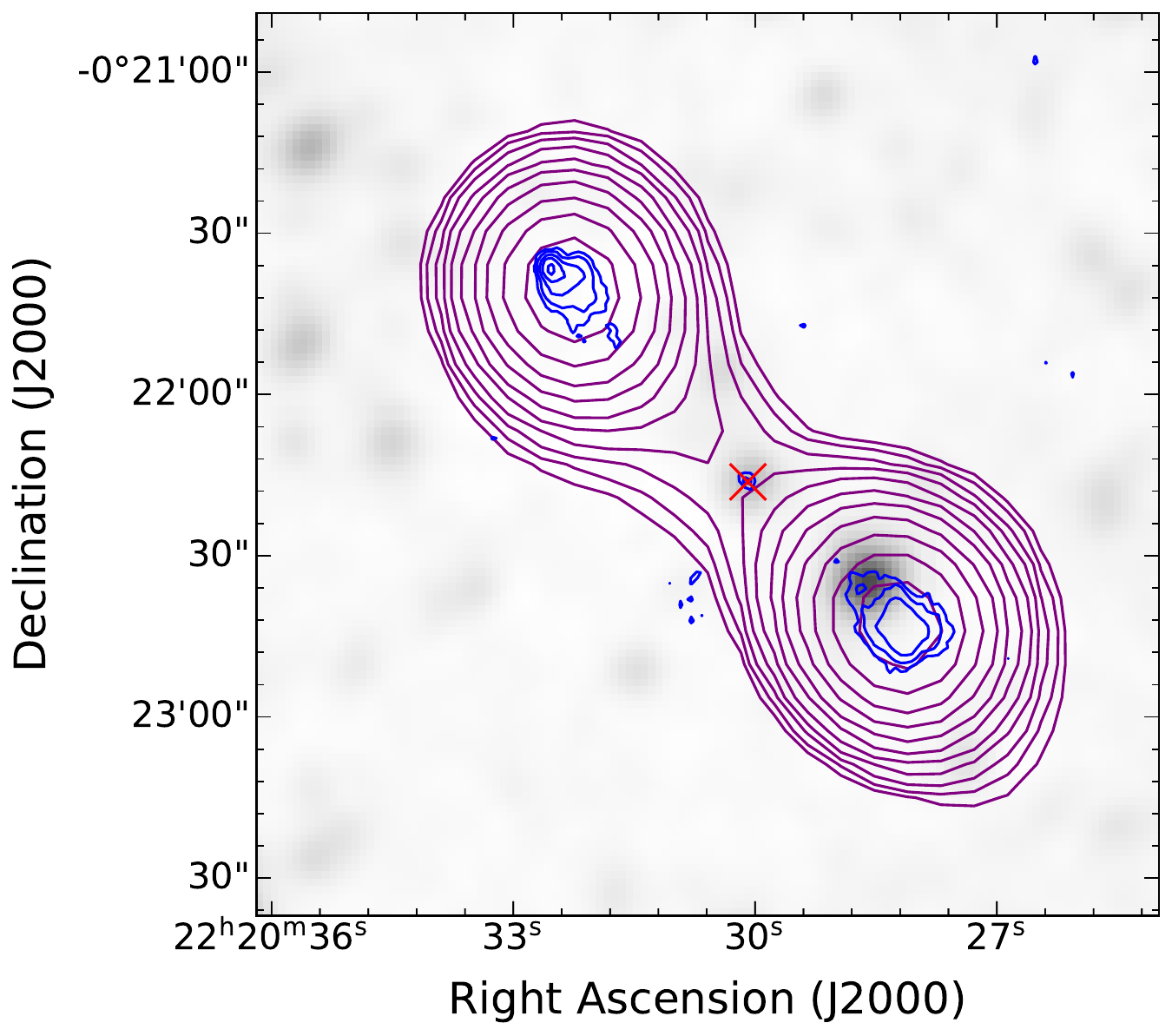} 
    \caption{Radio--optical overlay of 12 newly identified GRS candidates from our catalog. The images combine TGSS (purple) and VLASS (blue) radio continuum contours with AllWISE W1-band infrared‌ grayscale backgrounds. TGSS contour levels start at 3$\sigma_{\rm rms}$ (where $\sigma_{\rm rms}$ is the local RMS noise) and increase by factors of $\sqrt{2}$. VLASS contour levels start at 5$\sigma_{\rm rms}$ and increase by factors of 2. Red $\times$ symbols mark the host galaxy positions.
    \label{fig:grg_example}}
\end{figure*}

The catalog of GRS candidates from TGSS ADR1 is shown in Table~\ref{tab:radio_catalog}. Column (1) lists the short names of GRS candidates along with a flag, determined based on cross-matching results, indicating which of the sources in a set of catalogs (namely, \cite{2018ApJS..238....9K}, \cite{2020ApJS..247...53K}, \cite{2020A\string&A...635A...5D}, \cite{2020A\string&A...642A.153D}, \cite{2021ApJS..253...25K}, \cite{2021Galax...9...99A}, \cite{2022MNRAS.515.2032S}, \cite{2023A\string&A...672A.163O}, \cite{2024ApJS..273...30B}, \cite{2024A\string&A...686A..21S}, \cite{2024A&A...691A.185M}, \cite{2025A\string&A...699A.257A}, and \cite{2025ApJS..281...34M}). Among them, 69 GRSs from \cite{2018ApJS..238....9K}, 28 GRSs from \cite{2020A\string&A...635A...5D}, 140 GRSs from \cite{2020A\string&A...642A.153D}, 39 from \cite{2021ApJS..253...25K}, 5 GRSs from \cite{2021Galax...9...99A}, 4 from \cite{2022MNRAS.515.2032S}, 27 GRSs from \cite{2023A\string&A...672A.163O}, 26 GRSs from \cite{2024ApJS..273...30B}, 7 GRSs from \cite{2024A\string&A...686A..21S}, 982 GRSs from \cite{2024A&A...691A.185M}, one GRS from \cite{2025A\string&A...699A.257A}, and 25 GRSs from \cite{2025ApJS..281...34M}. There is substantial overlap among these twelve catalogs. After removing duplicates, we find that 1,029 GRSs were previously cataloged. Therefore, a total of 4,566 GRS candidates are newly identified in this study. Figure~\ref{fig:grg_example} displays the radio--optical overlays for 12 newly discovered GRS candidates. As shown in Figure~\ref{fig:grg_example}, the positions of the hosts and the VLASS radio cores in our GRS candidates are well coincident. This, together with the successful recovery of known GRSs from historical samples, further demonstrates the effectiveness of our methodology. Nevertheless, it is instructive to note that our catalog is not fully inclusive of the existing GRS catalog within the same region. Several factors account for this discrepancy. There are three primary reasons: first, the identification accuracy and completeness of FR-I and FR-II detection in our work are limited by the performance of the RGCMT model; second, although BT sources exist among previously known GRSs, they are not considered in our analysis; third, our GRS search method relies on the LR algorithm, whose accuracy in host galaxy identification depends critically on the input radio core coordinates and the depth of the photometric data. In this paper, a single radio core coordinate is adopted based on the geometric centroid along with AllWISE data, whereas the existing GRS catalogs in TGSS \citep[][]{2024ApJS..273...30B,2025ApJS..281...34M} use combined Panoramic Survey Telescope and Rapid Response System \citep[PanSTARRS;][]{2020ApJS..251....7F} and AllWISE data. Consequently, some sources remain unidentified as hosts due to the lack of corresponding counterparts in the AllWISE dataset.

\begin{deluxetable*}{lccccccccccccccc}
\digitalasset
\tablewidth{0pt}
\setlength{\tabcolsep}{3.3pt}
\tablecaption{A catalog of GRS candidates from TGSS ADR1. \label{tab:radio_catalog}}
\tablehead{
   \colhead{Short name} & \colhead{Name} & \colhead{R.A.}  & \colhead{Decl.} & \colhead{Typ.} & \colhead{$z$}  & \colhead{$F_{150}$} & \colhead{$F_{1400}$} & \colhead{$\alpha_{150}^{1400}$} & \colhead{BA}  & \colhead{LAS} & \colhead{LLS}  & \colhead{${\rm log}_{10}(P_{150})$} &\colhead{rmag} &  \colhead{$M_r$}&Class \\
    \colhead{(J2000)} & \colhead{(J2000)} & \colhead{(deg)}  & \colhead{(deg)} & \colhead{} & \colhead{}  & \colhead{(Jy)} & \colhead{(mJy)} & & \colhead{(deg)} & \colhead{(arcmin)} & \colhead{(Mpc)} & \colhead{(${\rm{W}}\,{{\rm Hz}^{ - 1}}$)} &(mag) & (mag)& 
}
\colnumbers
\startdata
J0000+1114$^{\rm m}$ &DESI J000051.10+111411.4 &0.21291 &11.23650 &Q &0.866$^{\rm S}$ &1.07 &260 & -0.63 & 17.7 & 3.23 &1.54 &27.5 & 17.09& -25.77 &II \\
J0000+1215$^{\rm d,k}$ & DESI LS J000043.36+121545.1 &0.18067 & 12.26253 & Gc & 0.914$^{\rm P}$ &1.58 &361 &-0.66 &2.1 &9.23 &4.46 &27.7 & 22.57 &-20.40 & II \\
J0000+1229 &WISEA J000024.09+122953.1 &0.10038 &12.49810&Gc &0.656$^{\rm P}$ &0.60&110 &-0.76 &2.3 &1.80 &0.78 &27.0 & 22.59 &-19.74 &II \\
J0000+2036 &DESI LS J000027.07+203602.3 &0.11279 &20.60064 &Gc &0.961$^P$ &0.49 &92 &-0.75 &13.7 &1.52 &0.74 &27.3 & 23.01 &-20.05 &II \\
J0000$-$0721 &DESI LS J000039.82$-$072129.5 &0.16592 &-7.35819 &Qc &0.474$^{\rm P}$ &0.46 &133 &-0.55 &20.7 &2.68 &0.99 &26.5 &13.6 &-28.06 &II \\
J0000$-$2450 &DESI LS J000019.00-245049.7 &0.07917 &-24.84714 &Gc &0.963$^{\rm P}$ &0.18 &50 &-0.58 &3.8 &2.12 &1.04 &26.8 & 22.26 &-20.80&II \\
J0000$-$3410 &DESI LS J000016.30-341022.4 &0.06792 &-34.17289 &Gc &0.724$^{\rm P}$ &0.68 &130 &-0.74 &6.2 &2.35 &1.05 &27.1 &20.54 &-21.99 &II\\
J0000$-$3424 &DESI LS J000042.11$-$342403.4 &0.17546 &-34.40094 &Gc &1.214$^{\rm P}$ &2.99 &398 &-0.9 &35.6 &2.23 &1.14 &28.4 &21.37&-22.06 &II \\
J0001+0820$^{\rm m}$ &DESI J000114.98+082029.1 &0.31241 &8.34142 &Q &0.544$^{\rm S}$ &2.10 &405 &-0.74 &28.9 &2.92 &1.15 &27.3 & 18.97&-22.98&II \\
J0001+0936 &DESI LS J000126.00+093631.9 &0.35833 &9.60886 &Gc &1.036$^{\rm P}$ &0.70 &142 &-0.71 &1.2 &1.92 &0.95 &27.5 &23.89&-19.30&II \\
\multicolumn{1}{c}{...} &... & ... & ... & ... & ... & ... & ... & ... & ... & ... & ... & ... & ...& ... \\
\enddata
\tablecomments{Column (1) short name (JHHMM+DDMM); flags `a', `b', `c', `d', `e', `f', `g', `h', `i', `j', `k', `l', and `m' in the short names indicate the sources present in \cite{2018ApJS..238....9K}, \cite{2020ApJS..247...53K}, \cite{2020A\string&A...635A...5D}, \cite{2020A\string&A...642A.153D}, \cite{2021ApJS..253...25K}, \cite{2021Galax...9...99A}, \cite{2022MNRAS.515.2032S}, \cite{2023A\string&A...672A.163O}, \cite{2024ApJS..273...30B}, \cite{2024A\string&A...686A..21S}, \cite{2024A\string&A...691A.185M}, \cite{2025A\string&A...699A.257A}, and \cite{2025ApJS..281...34M}, respectively. Column (2): full name derived either from the host galaxy name following the redshift-reference catalog or imaging survey in J2000. Columns (3) and (4): coordinates corresponding to the full names in deg. Column (5): spectral type of the host galaxy, where G=galaxy, Gc=galaxy candidate, Q=QSO, Qc=QSO candidate. Column (6): redshift ($z$) with a flag indicating whether it is spectroscopic (`S') or photometric (`P'). Column (7): integrated flux densities at 150 MHz in Jy. Column (8): integrated flux densities at 1400 MHz in mJy. Column (9): spectral index between 150 MHz and 1400 MHz. Column (10): the bending angle (BA) in deg. Column (11): the largest angular size (LAS) in arcmin. Column (12): the largest (projected) linear size (LLS) in Mpc, which were rounded to two decimal places for table consistency; the precision is limited by the $\sim25^{\prime\prime}$ angular resolution and redshift uncertainties. Column (13): the logarithm radio power in ${\rm{W}}\,{{\rm Hz}^{ - 1}}$ at 150 MHz. Column (14):$r$-band absolute magnitudes in mag. Column (15): Fanaroff-Riley (FR) type, `I' indicates FR I, `II' represents FR II, and `I/II' is source shows characteristics of both FR I and FR II. Table \ref{tab:radio_catalog} is published in its entirety in the
electronic edition of \href{https://doi.org/10.3847/1538-4365/ae96ae}{\textit{the Astrophysical Journal Supplements}}. A portion is shown here for guidance regarding its form and content.} 

\end{deluxetable*}

\begin{figure*}[ht!]
\centering
\includegraphics[scale=0.9]{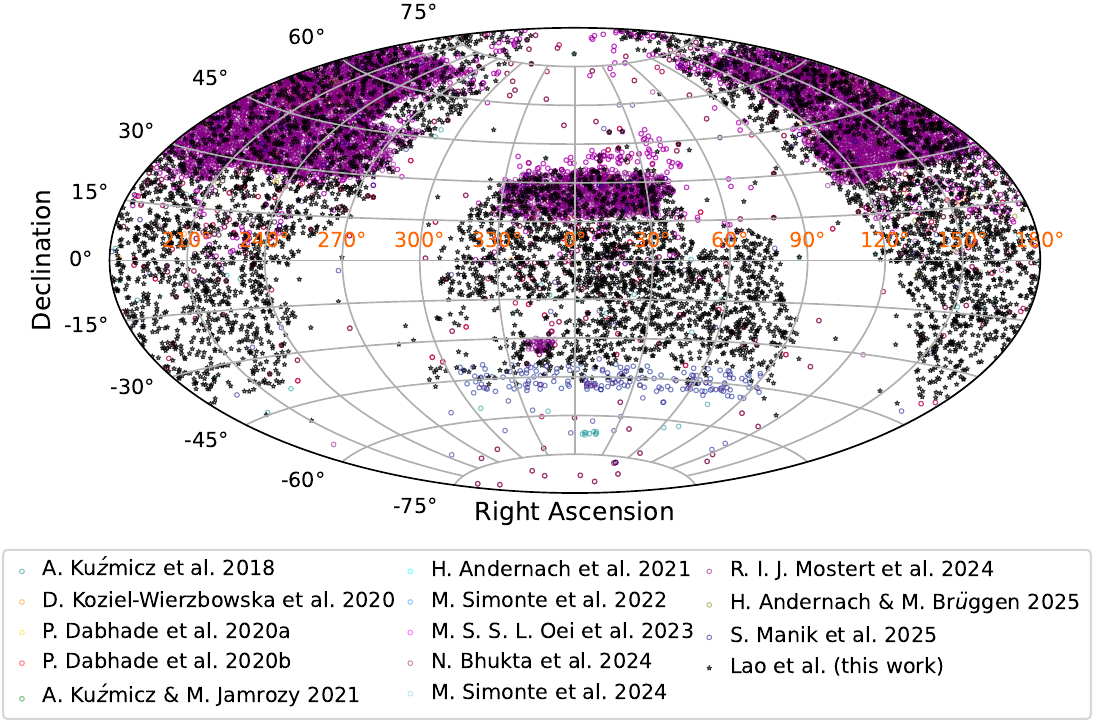}
\caption{All-sky distribution of GRS candidates from this work (black pentagrams) and from 13 large samples (colored circles): \cite{2018ApJS..238....9K} (dark cyan), \cite{2020ApJS..247...53K} (dark orange), \cite{2020A&A...635A...5D} (gold), \cite{2020A&A...642A.153D} (red), \cite{2021ApJS..253...25K} (green), \cite{2021Galax...9...99A} (cyan), \cite{2022MNRAS.515.2032S} (dodger blue), \cite{2023A&A...672A.163O} (magenta), \cite{2024ApJS..273...30B} (brown), \cite{2024A&A...686A..21S} (sky blue), \cite{2024A&A...691A.185M} (purple), \cite{2025A&A...699A.257A} (olive), \cite{2025ApJS..281...34M} (navy).}
\label{fig:sky_map}
\end{figure*}

Column (2) provides the full names of GRS candidates. These names adopt the host galaxy naming conventions from the redshift-reference catalog or imaging survey that provided the spectroscopic or photometric redshift. Columns (3) and (4) list the J2000 epoch Right Ascension (R.A.) and Declination (Dec.) for each GRS candidate in decimal degrees. Figure~\ref{fig:sky_map} shows the all-sky projection distribution of GRS candidates in our catalog compared with existing GRS catalogs, demonstrating that our sample not only reproduces known sources from previous catalogs but also provides substantial additions in previously uncovered regions of the sky. Column (5) presents the spectra types of host galaxies for our GRS candidates, where `G' indicates galaxy, `Gc' represents galaxy candidate, `Q' is the QSO, and `Qc' is the QSO candidate. The types of hosts were determined through a three-step procedure. First, within a 1$''$ radius centered on each host's coordinates, we sequentially cross-matched against the following catalogs: the `specobjall' table from SDSS DR17, the `agngal' table from the DESI DR1 Value-Added Catalogs, the galaxy catalog of DESI LS presented in \citet{2023MNRAS.526.4768W}, the QSO catalog in \citet{2023OJAp....6E..49F}, and the QSO catalog from the Value-Added Catalog of the nineteenth Data Release of SDSS \citet[SDSS DR19;][]{2025arXiv250707093S}. The types of matched hosts were assigned based on the corresponding classifications in these catalogs. Second, for those hosts that remained unmatched in the first step, we continued the search within the same matching radius against the QSO candidate catalog from the DESI LS DR9 \citep{2022RAA....22i5021H}, and hosts matched at this stage were classified as QSOs. Finally, for those hosts whose types could not be determined through the existing catalogs, we adopted the star-galaxy-QSO classification method presented in \citet{2025ApJS..276...46L}. Hosts identified as QSOs and galaxies by this classification method were designated as QSO candidates and galaxy candidates, respectively, while the remaining hosts were classified as undetermined. In our GRS candidate catalog, there are 957 galaxies, 3,253 galaxy candidates, 638 QSOs, and 392 QSO candidates. The remaining 355 GRSs have uncertain types due to the lack of definitive classification information for their host galaxies. Therefore, 4,210 are GRGs, and 1,030 are GRQs.

Column (6) lists the redshift of the GRS host. In our GRS candidate catalog, 1,431 sources have spectroscopic redshifts with flag `S', and 4,164 sources have photometric redshifts with flag `P'. Among the spectroscopic redshifts, 862 were extracted from the `agngal' table of the DESI DR1 Value-Added Catalogs, 423 from the `specobjall' table of SDSS DR17, 3 from the spectra and redshifts catalog of the 2dFGRS, 20 from the final redshift release of the 6dFGS, 5 from the final data release of the WiggleZ Dark Energy Survey, and 118 from the NED database. Among the photometric redshifts, 3,951 were obtained from the `tractor' table of the DESI LS DR10, and 213 from the NED database. The redshifts of our GRS candidates range from $z = 0.056$ to at least $z = 2.385$ for sources with secure measurements, with a median of $z = 0.875$, while an additional 27 candidates tentatively extend the redshift boundary up to $z = 4.254$.

\begin{figure}[ht!]
\centering
\includegraphics[scale=0.5]{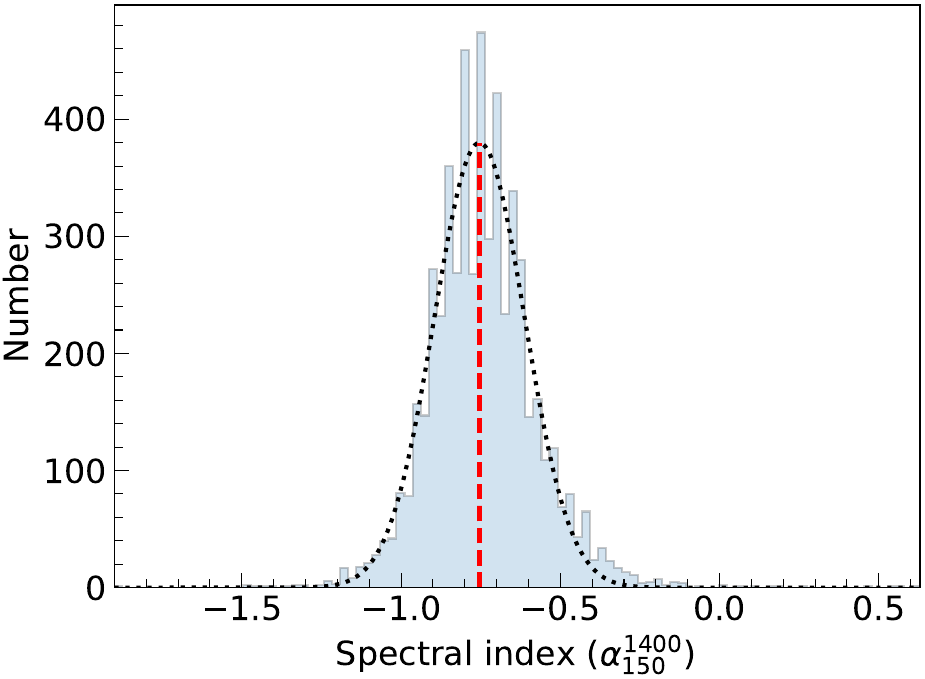}
\caption{Histogram showing the distribution of two-point spectral indices ($\alpha_{\rm 150}^{\rm 1400}$) for our GRS candidates, computed from integrated flux densities at 150 MHz and 1400 MHz (hence representing each source in its entirety, not just its lobes or core). The black dotted curve represents a Gaussian fit to the distribution (mean=-0.75$\pm$0.01, standard deviation=0.14), and the red dashed line marks the mean value.  
\label{fig:spi}}
\end{figure}

Columns (7) and (8) provide the integrated flux densities for GRS candidates at 150 MHz and 1400 MHz, respectively. The 150 MHz integrated flux densities of our GRS candidates were calculated from TGSS ADR1 image data, while the 1400 MHz integrated flux densities were derived from NVSS maps. All GRS candidates have calculated TGSS integrated flux densities; 5,519 GRS candidates have NVSS integrated flux densities, as the remaining sources lack NVSS cutout maps from SkyView and thus have no results. Our GRS candidates have 150 MHz integrated flux densities ranging from $\sim$20~mJy to $\sim$151~Jy. Column (9) lists the two-point spectral indices ($\alpha^{\rm 1400}_{\rm 150}$) between the TGSS and NVSS integrated flux densities. A total of 5,519 GRS candidates were derived for $\alpha^{\rm 1400}_{\rm 150}$. Figure~\ref{fig:spi} shows the spectral index distribution for our GRS candidates, with a median spectral index of $\alpha^{\rm 1400}_{\rm 150} = -0.75$ and a mean spectral index of $\alpha^{\rm 1400}_{\rm 150} = -0.74$. This distribution shows a slightly steeper spectrum compared to the canonical value of $-$0.7 for extended synchrotron emission in radio galaxies \citep[e.g.,][]{2010MNRAS.408.2261K}. Among the catalog, 418 GRS candidates (7.6\%) are found to have a flat radio spectrum ($\alpha^{\rm 1400}_{\rm 150} > -0.5$), 5079 GRS candidates (92.0\%) exhibit steep spectra ($-1.2 \lesssim \alpha^{\rm 1400}_{\rm 150} \lesssim -0.5$), and 22 GRS candidates (0.4\%) are ultra-steep-spectrum sources ($\alpha^{\rm 1400}_{\rm 150} < -1.2$).


\begin{figure}[ht!]
\centering
\includegraphics[scale=0.5]{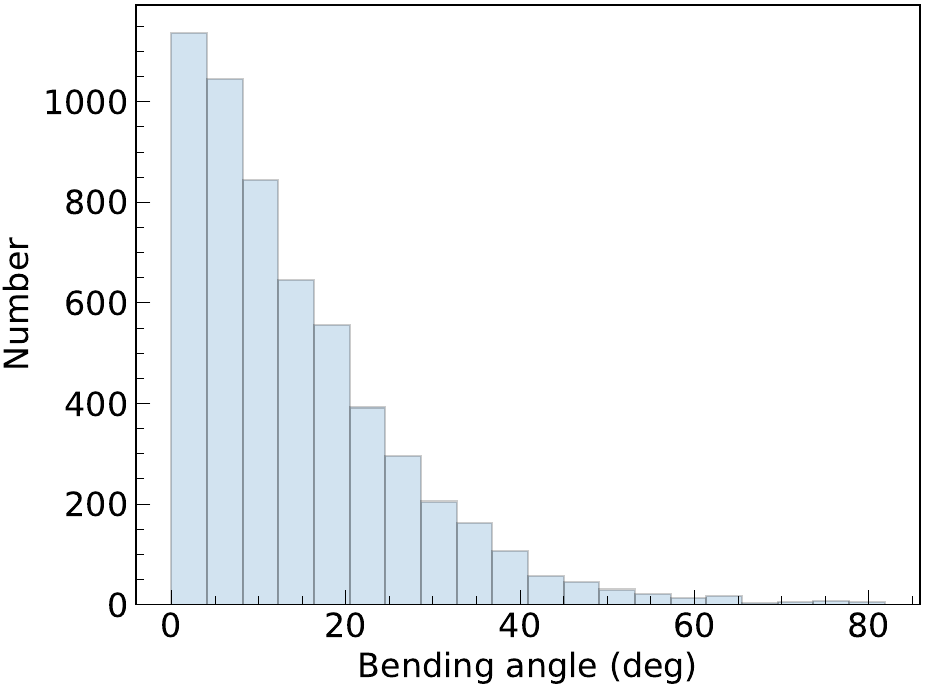}
\caption{Histogram showing the distribution of bending angle (BA) for our GRS candidates.  
\label{fig:BA}}
\end{figure}

\begin{figure}[ht!]
\centering
\includegraphics[scale=0.5]{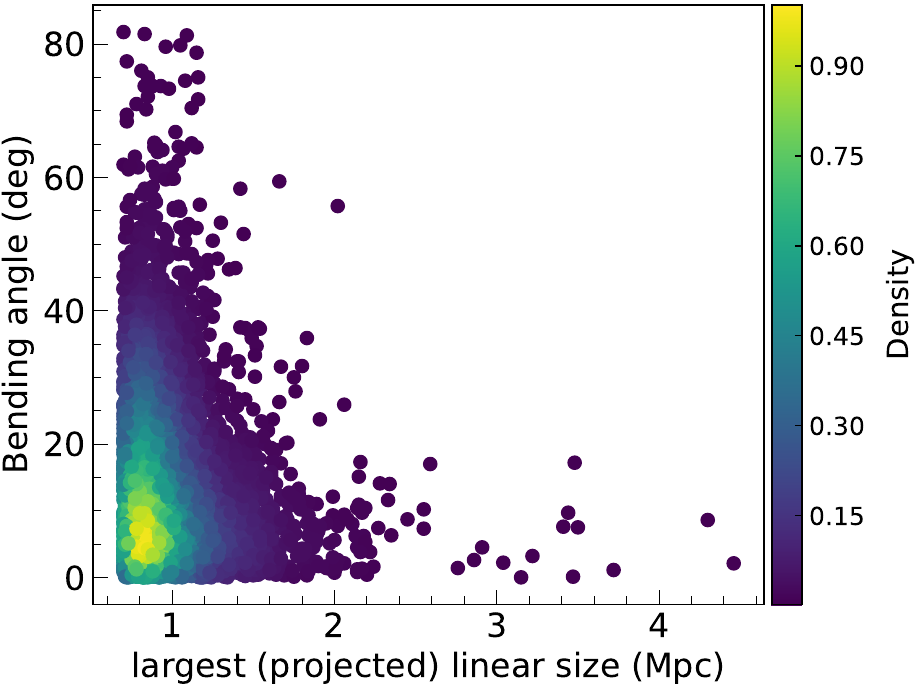}
\caption{Density-scatter plot of the bending angle (BA) as a function of the largest (projected) linear size (LLS) in units of Mpc. The color bar indicates the normalized source density. 
\label{fig:lls_BA}}
\end{figure}

Column (10) lists the bending angle (BA) of all GRS candidates. For the BA of each GRS candidate, we first calculated its opening angle (OA) based on the mask provided by the RGCMT model, following the method described in \cite{2025ApJS..276...46L}. The BA was then derived by subtracting the OA from 180 degrees. Figure~\ref{fig:BA} shows the distribution of BA, with a mean and median BA of 14.3$^\circ$ and 11.0$^\circ$. For the vast majority (96\%) of GRS candidates, the BA lies between 0$^\circ$ and 40$^\circ$. This is consistent with the view presented in \cite{2025A&A...699A.257A} that GRSs have relatively small BA values. Columns (11) and (12) present the LAS and LLS of our GRS candidates. The LAS have a range of 1.35$'$ to 15.42$'$, and the LLS have a value from $\sim$0.7 Mpc to $\sim$4.5 Mpc. The relationship between the BA and the LLS for our GRS candidates is presented in Figure~\ref{fig:lls_BA}. The distribution exhibits a characteristic ``L-shaped" density profile, with a primary concentration of sources situated at LLS$<$1.5 Mpc and BA$<$30$^\circ$. This high-density region represents the typical population of relatively straight radio galaxies, consistent with those residing in lower-density environments \citep[e.g.,][]{2015MNRAS.449..955M,2020A&A...638A..48S} or those aligned close to the plane of the sky \citep{2026A&A...706A.310M}. A notable trend observed in Figure~\ref{fig:lls_BA} is the absence of sources with both a large LLS ($>$2.5 Mpc) and a big BA ($>$20$^\circ$). This deficit is physically significant; GRSs generally extend into the tenuous intergalactic medium (IGM), where the ram pressure, primary driver of jet bending, is significantly lower than in the dense cores of galaxy clusters. Consequently, these large-scale structures are less susceptible to the large-scale bending typically seen in BT sources. Furthermore, the scatter toward higher BAs at smaller LLSs may be partially attributed to projection effects. Sources with significant intrinsic bending that are oriented close to the line of sight will appear foreshortened in LLS, while their apparent BA is geometrically exaggerated. However, the high density of moderately bent sources at about 1 Mpc suggests that environmental interactions within the intracluster medium (ICM) play a dominant role in shaping the morphology of these radio jets. 



\begin{figure*}
    \centering
    \includegraphics[scale=0.5]{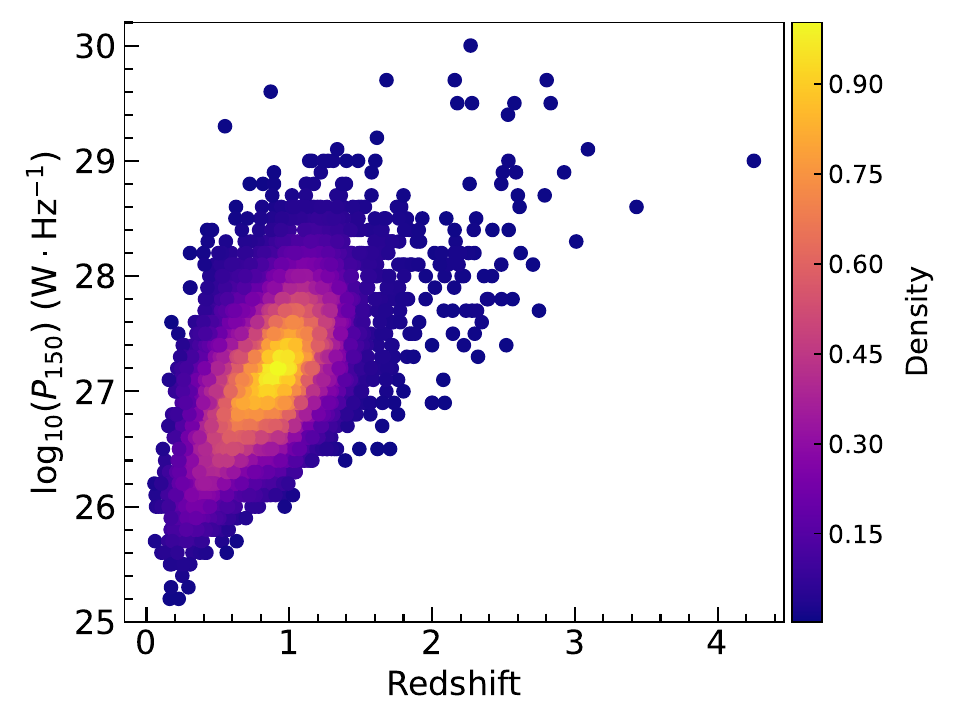}
    \hspace{2mm}
    \includegraphics[scale=0.5]{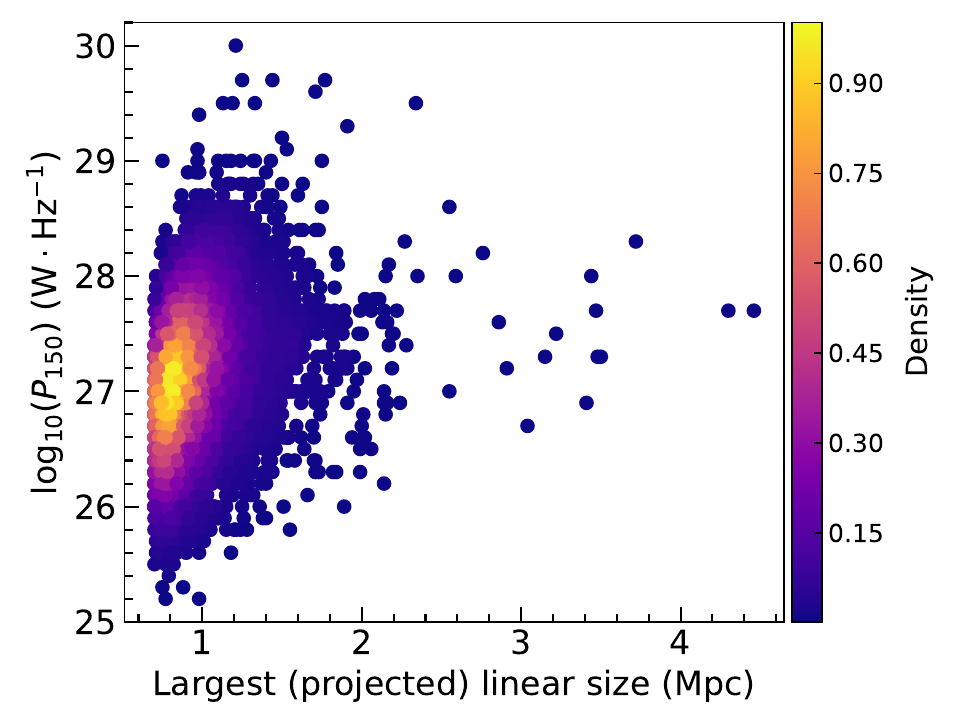}
    \caption{Left: Density-scatter plot of the logarithm of radio power at 150 MHz (${\rm log}_{10}(P_{150})$) as a function of redshift. Right: ($P$--$D$ diagram) Density-scatter plot of the ${\rm log}_{10}(P_{150})$ as a function of the largest (projected) linear size (LLS) in units of Mpc.
    \label{fig:Power_z_lls}}
\end{figure*}

The 150 MHz radio power ($P_{\rm 150}$) of our GRS candidates, listed in Column (13), were calculated using the standard formula \citep[e.g.,][]{2009MNRAS.392..617D}: 
\begin{equation}
{P_{\rm 150}} = \frac{{4{\rm \pi}{F_{\rm 150}} {{({D_{L}})}^2}}}{{{{(1 + z)}^{(1 + \alpha^{\rm 1400}_{\rm 150})}}}} ,
	\label{eq:power}
\end{equation}
where $D_L$ is the luminosity distance to the GRS candidate, $F_{\rm 150}$ is the integrated flux density at 150~MHz, $z$ is the redshift of the source, and $\alpha^{\rm 1400}_{\rm 150}$ is the spectral index between 1400~MHz and 150~MHz. The left panel of Figure~\ref{fig:Power_z_lls} shows the distribution of the logarithm of radio power at 150 MHz (${\rm log}_{10}(P_{150})$) as a function of redshift. In our sample, ${\rm log}_{10}(P_{150})$ values range from $\sim$25.2 to $\sim$30.0 ${\rm W\,Hz^{-1}}$, with mean and median values of 27.2 and 27.1 ${\rm W\,Hz^{-1}}$, respectively. 

Plotting radio power as a function of LLS in a $P$--$D$ diagram has proven to be an effective way to visualize the relationship between these two parameters \citep{1982IAUS...97...21B}. We present the $P$--$D$ diagram of our GRS candidates in the right plane of Figure~\ref{fig:Power_z_lls}, where the radio power at 150 MHz (${\rm log}_{10}(P_{150})$) is plotted against the LLS. The color-coding represents the point density, revealing that the bulk of the population is concentrated at LLS$<$1.5 Mpc and 26$<{\rm log}_{10}(P_{150})<$28 ${\rm W}\,\,{\rm Hz}^{-1}$. Consistent with the canonical dynamical and luminosity evolution models of extragalactic radio sources \citep[KA97, KDA97;][]{1997MNRAS.286..215K, 1997MNRAS.292..723K}, as well as the recent observational findings of \citet{2025A&A...699A.257A} that GRSs, being old radio galaxies, are less powerful than smaller radio sources, there is a conspicuous dearth of sources in the upper-right corner of the $P$--$D$ diagram. This absence suggests that very powerful radio sources (${\rm log}_{10}(P_{150})>$28.5 ${\rm W}\,\,{\rm Hz}^{-1}$) rarely maintain such high luminosity as they expand into the GRS regime, possibly due to adiabatic expansion and synchrotron losses. Conversely, the scarcity of sources in the bottom-right corner (low power, large LLS) is primarily driven by selection effects; as the surface brightness of a source decreases quadratically with increasing LLS, those with low integrated flux density fall below the survey's S/N detection threshold \citep[e.g.,][]{2023A&A...672A.163O,2024A&A...691A.185M,2025A&A...699A.257A}. However, several high-luminosity outliers exist with $\log_{10}(P_{150}) > 28.5$~W~Hz$^{-1}$ and LLS exceeding 1~Mpc, occupying precisely the region where self-similar models predict a dearth of sources (the upper-right corner of the $P$--$D$ diagram). These objects challenge the same evolutionary frameworks that successfully explain the overall population: the KDA97 model predicts that, owing to catastrophic radiative and adiabatic energy losses, radio power should decline steeply with increasing size. Maintaining $\log_{10}(P_{150}) > 28.5$~W~Hz$^{-1}$ at $\text{LLS} > 1$~Mpc would therefore require either extraordinarily high initial jet powers $> 10^{46}$~erg~s$^{-1}$ (an order of magnitude above typical high-excitation radio galaxies) or continuous in-situ particle re-acceleration. Without such additional mechanisms, these outliers lie more than $3\sigma$ above the expected KDA97 $P$--$D$ evolutionary track, placing them in direct tension with the simplest no-re-acceleration, constant-power models. These ``over-luminous" GRG candidates represent critical cases for constraining the duty cycle and energy injection rates of AGN jets.

\begin{figure}[ht!]
\centering
\includegraphics[scale=0.5]{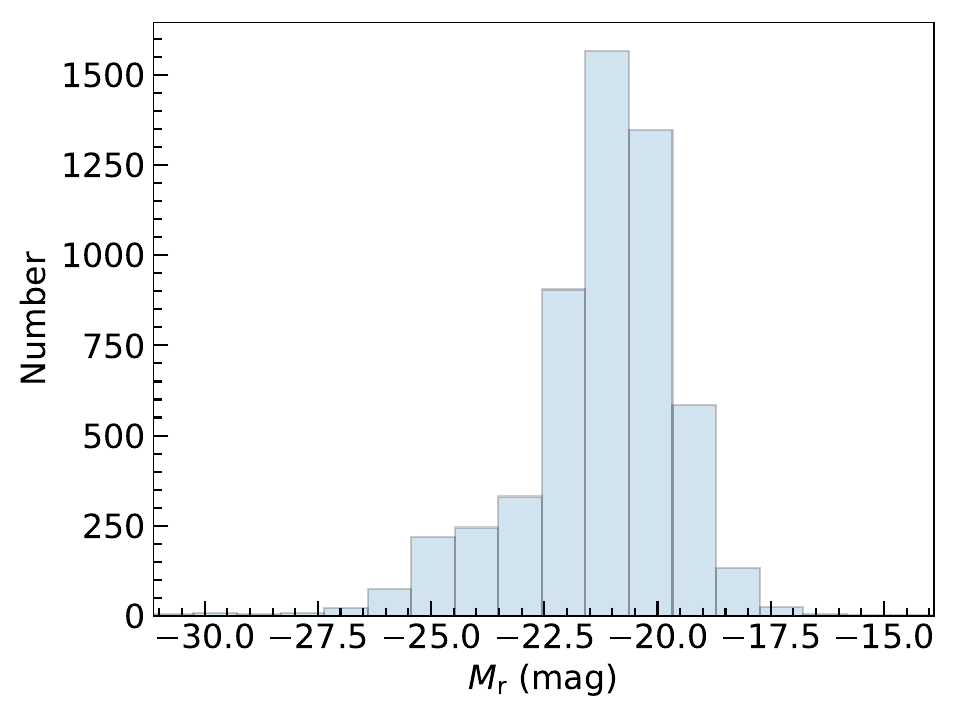}
\caption{Distribution of the $r$-band absolute magnitudes ($M_r$) for 5,486 GRS candidates.  
\label{fig:Mr}}
\end{figure}

Columns (14) and (15) provide the $r$-band apparent magnitude (rmag) and absolute magnitude ($M_r$) of the hosts for our GRS candidates. The apparent magnitudes were directly retrieved from the `photoplate' table of SDSS DR17 and the main tractor photometry catalog of DESI LS DR10. Specifically, SDSS DR17 provides model magnitudes (de Vaucouleurs/exponential profile fits), while DESI LS DR10 provides Tractor magnitudes derived from the best-fit among five morphological models: S\'ersic (SER), de Vaucouleurs (DEV), exponential (EXP), round exponential (REX), or point spread function (PSF). Following \cite{2024AJ....168...58D}, the $M_r$ values for all GRS candidates were computed as:
\begin{equation}
M_r = rmag - 5 \log_{10}(D_L) - 25 - K_{\rm corr} - E_{\rm corr},
	\label{eq:Mr}
\end{equation}
where the $K$-correction is defined as $K_{\rm corr}=-2.5\log_{10}(1+z_{\rm ref})$ with a reference redshift of $z_{\rm ref}=0.1$ \citep{2024AJ....168...58D}, and the evolutionary correction is given by $E_{\rm corr} = -Q_0(z - z_{\rm ref})$ \citep{2014MNRAS.445.2125M}, adopting $Q_0 = 0.97$. In the catalog, $M_r$ has been calculated for a total of 5,486 sources. Its distribution is shown in Figure~\ref{fig:Mr}, with a mean and median of $M_r=-21.26$ and $M_r=-21.02$ mag, respectively. Most (99.2\%) of our GRS candidates have $M_r$ values lying in the range $-27 \lesssim M_r \lesssim -17$ mag. 

\begin{figure}[ht!]
\centering
\includegraphics[scale=0.5]{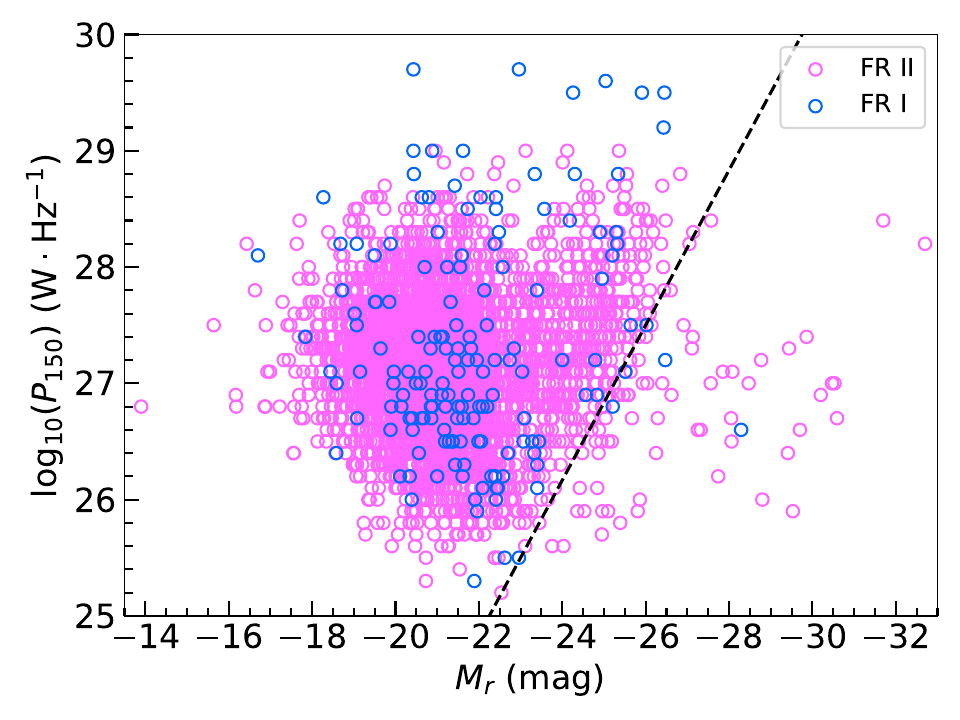}
\caption{The 150 MHz radio power (${\rm log}_{10}(P_{150})$) vs. $r$-band absolute magnitude ($M_r$) for the FR I and FR II sources in our GRS candidate catalog. The power division line (the black dashed line) is from \citet{1996AJ....112....9L}.  
\label{fig:P_Mr}}
\end{figure}

Our catalog contains sources that showed characteristics of both FR I and FR II. Because the RGCMT model did not perform radio morphology subdivision, these sources were included as a subclass with hybrid FR I and FR II features. We therefore adopted visual inspection to reclassify our sources and identify those exhibiting characteristics of both FR I and FR II. Column (16) lists the radio morphologies of our GRS candidates, where `I' and `II' indicate the source exhibits FR I and FR II morphologies, respectively, and `I/II' represents a source that exhibits both morphologies. In the catalog, 4,656 sources (83.2\%) are FR II, 176 sources (3.2\%) are FR I, and 763 sources (13.6\%) have both FR I and FR II morphologies. This agrees with the points of \cite{2020A&A...635A...5D} that most GRSs exhibit an FR II morphology, while few GRSs exhibit an FR I morphology, and a small fraction exhibit a hybrid morphology. Figure~\ref{fig:P_Mr} shows the 150 MHz radio power (${\rm log}_{10}(P{150})$) against the $M_r$ for the FR I and FR II sources. Most FR IIs lie above and to the left of the power division line (the black dashed line in Figure~\ref{fig:P_Mr}), with only a few located below and to the right. This indicates that the FR II sources show evidence of following the power division used to identify FR IIs, as in \cite{1974MNRAS.167P..31F,1996AJ....112....9L}. Only four FR I sources are located below and to the right of the power division line, while almost all FR I sources lie above and to the left. This agrees with our previous work in \citet{2026ApJS..283...11L} that the power division may not strictly identify the FR I sources.

\subsection{Pipeline Performance Evaluation} 
\begin{figure*}
    \centering
    \includegraphics[scale=0.55]{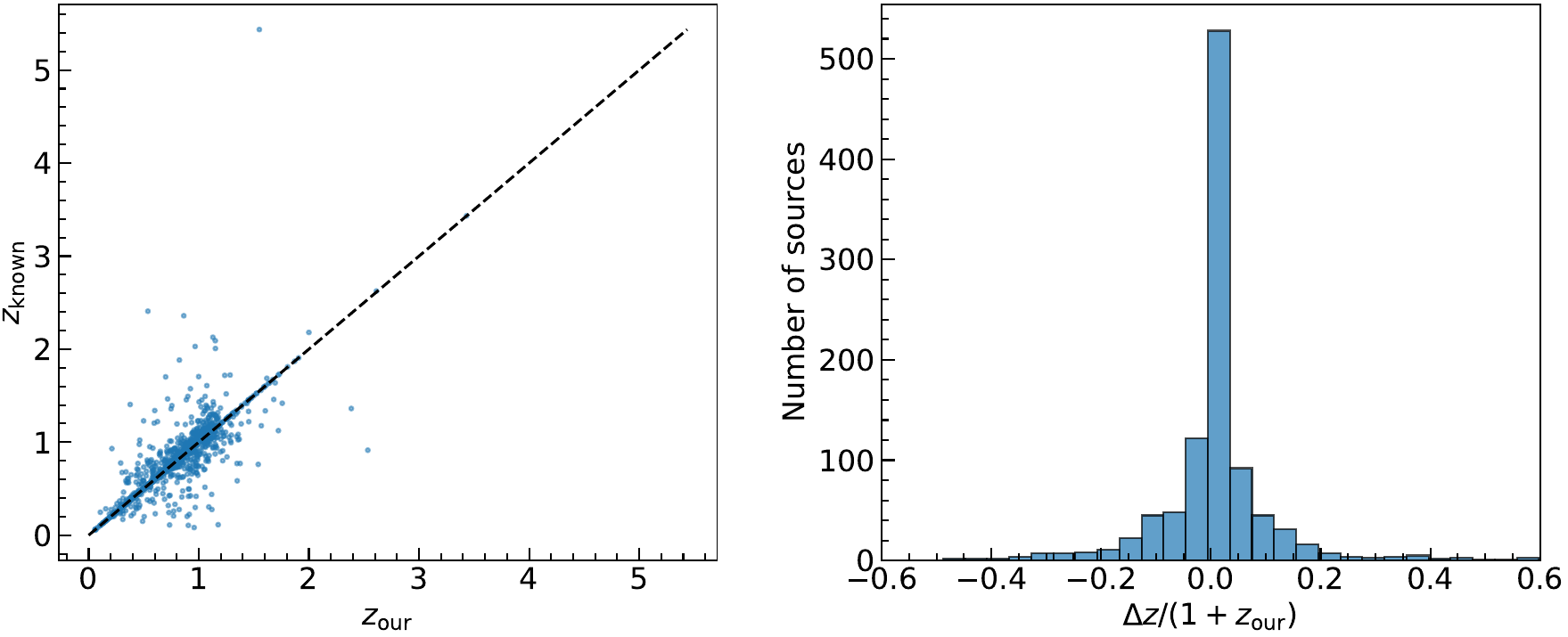}
    \includegraphics[scale=0.55]{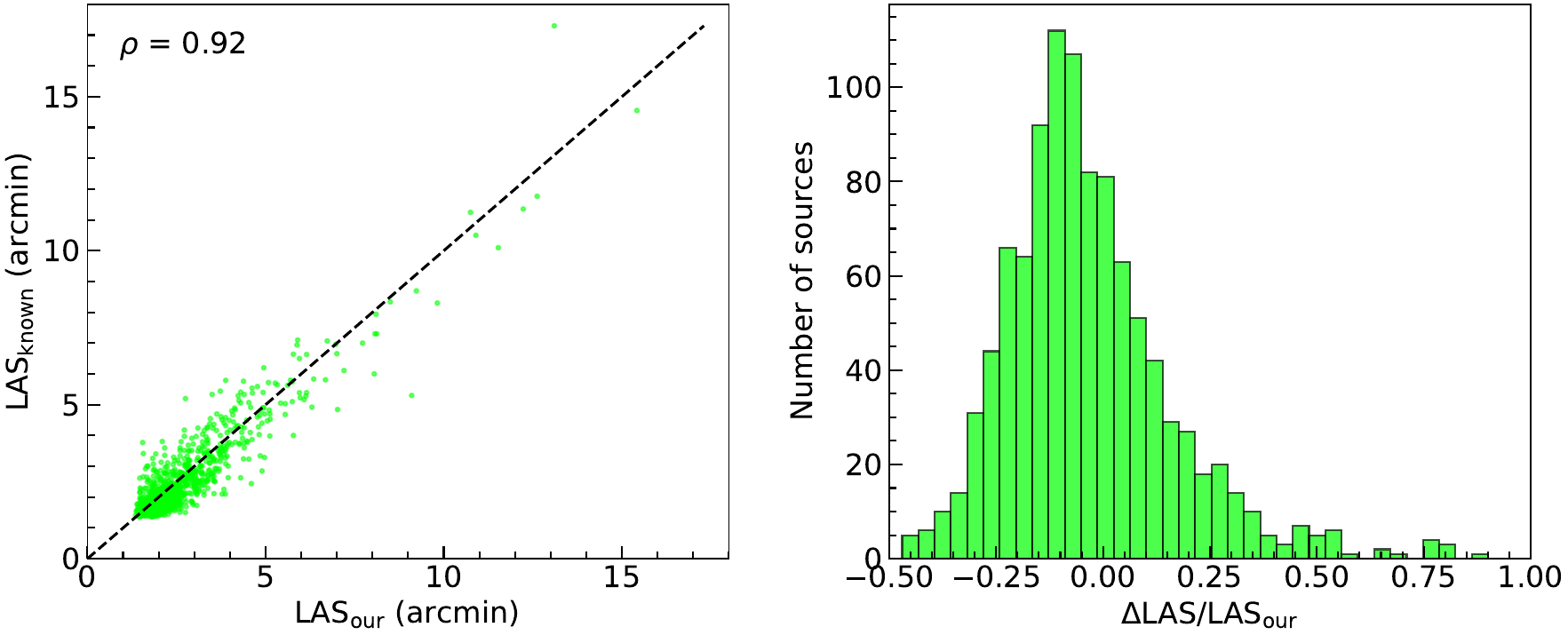}
    \includegraphics[scale=0.55]{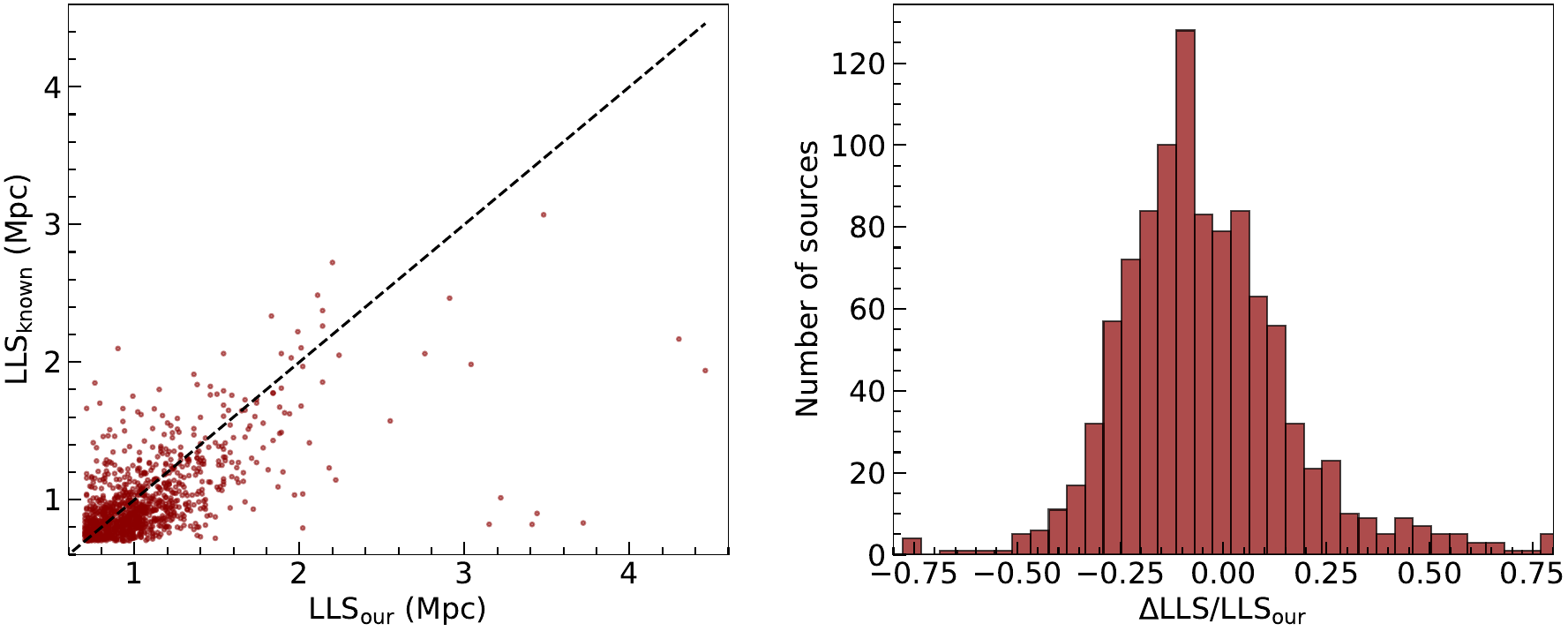}
    \caption{Top left: Redshift $z_{\rm our}$ of our GRS candidates versus $z_{\rm known}$ of matched sources from previous GRS catalogs. The dashed line indicates the 1:1 relation. Top right: Histogram of the normalized redshift residual $\Delta z/(1+z_{\rm our})$ (with $\Delta z = z_\mathrm{known} - z_\mathrm{our}$) for the full set of matched sources. Middle left: Comparison of LAS between our GRS candidates ($\mathrm{LAS}_{\rm our}$) and matched sources from previous catalogs ($\mathrm{LAS}_{\rm known}$). The dashed line marks the 1:1 relation. The Pearson correlation coefficient is $\rho = 0.92$. Middle right: Distribution of the normalized LAS residual, $\Delta \mathrm{LAS} / \mathrm{LAS}_{\rm our}$ (with $\Delta {\rm LAS} = {\rm LAS}_\mathrm{known} - {\rm LAS}_\mathrm{our}$), for all matched sources. Bottom left: Comparison of LLS values derived from our method (LLS$_{\rm our}$) versus those from previous work (LLS$_{\rm known}$). The dashed line represents 1:1 perfect agreement. Bottom right: Distribution of the relative LLS difference, $\Delta \mathrm{LLS} / \mathrm{LLS}_{\rm our}$ (with $\Delta {\rm LLS} = {\rm LLS}_\mathrm{known} - {\rm LLS}_\mathrm{our}$), for the matched sources.
    \label{fig:z_las_lls_anal}}
\end{figure*}

This section evaluates our pipeline's performance and the validity of its candidate discoveries. The analysis focuses on two complementary aspects: (1) for GRS candidates with matches in existing catalogs, we compare their redshift, LAS, and LLS with those from previous catalogs; (2) for unmatched candidates (potential new discoveries), we perform visual inspection and incorporate redshift uncertainties for further assessment. As described in Section \ref{sec:catalog}, there are 1,029 GRS candidates in previous catalogs. For a candidate appearing in two or more previous catalogs, we compute the average of its redshift, LAS, and LLS from those catalogs and compare these averages with those of our GRS candidate. If it appears in only one previous catalog, we directly use its redshift, LAS, and LLS for comparison with our GRS candidate. Figure~\ref{fig:z_las_lls_anal} shows the comparison of redshift, LAS, and LLS for the matched GRSs.

The top two panels of Figure~\ref{fig:z_las_lls_anal} present the redshift comparison results for the full sample of 1,029 cross-matched GRS candidates collected from pre-existing catalogs. The left panel plots our retrieved redshift ($z_\mathrm{our}$) against the literature redshift ($z_\mathrm{known}$) adopted from publicly available GRS catalogs, with the black dashed line marking the 1:1 perfect consistency reference. Among the entire matched sample, 1,012 sources, accounting for 98.3\% of the total, have their normalized redshift difference $\Delta z/(1+z_\mathrm{our})$ (with $\Delta z = z_\mathrm{known} - z_\mathrm{our}$) distributed within the interval of $-0.4$ to $+0.4$, as clearly shown by the sharp central peak in the histogram on the right panel. Just 1.6\% of sources lie in the extended tails outside this range, across the full axis span of $-0.6$ to $+0.6$. 98.3\% of the data points are clustered around the 1:1 reference line, confirming an excellent overall agreement between the redshift values we retrieved and those previously reported in the literature. Only a few sources show obvious systematic deviations. The most prominent outlier corresponds to the GRS candidate J1557+3647: our extraction gives a redshift of $z=1.550$ for this source, while its counterpart in the catalog of \cite{2024A&A...691A.185M} has been assigned a much higher redshift of $z=5.435$. After re-examining the multi-band spectroscopic data and redshift fitting pipeline for this source, we conclude that the redshift value reported in \cite{2024A&A...691A.185M} is probably overestimated. This source corresponds to the isolated outlier point at the top left corner of the left-hand scatter plot.

The middle two panels of Figure~\ref{fig:z_las_lls_anal} present the LAS comparison results for the full sample of cross-matched GRS candidates collected from pre-existing public catalogs. The left scatter panel plots our pipeline-retrieved LAS ($\mathrm{LAS_{our}}$) against the literature LAS ($\mathrm{LAS_{known}}$) extracted from published GRS catalogs, with the black dashed line marking the 1:1 perfect consistency reference. The calculated Pearson correlation coefficient between the two sets of LAS values reaches $\rho = 0.92$, demonstrating a strong positive linear correlation. Among the entire matched sample, 1005 sources, accounting for 97.6\% of the total, have their relative LAS difference $\Delta \mathrm{LAS}/\mathrm{LAS_{our}}$ (defined as $\Delta \mathrm{LAS} = \mathrm{LAS_{known}} - \mathrm{LAS_{our}}$) distributed within the interval of $-0.5$ to $+0.5$, forming a prominent sharp central peak in the normalized difference histogram on the right panel. Only 2.4\% of sources extend into the long tails across the full axis span up to $\pm1.0$, and 97.6\% of the data points are clustered along the 1:1 reference line, which robustly confirms that the LAS values output by our measurement pipeline show excellent overall consistency with the LAS measurements of the same matched GRSs in pre-existing catalogs. Only a small number of sources exhibit noticeable deviations far from the 1:1 reference line. The most prominent outlier at the top region of the scatter plot corresponds to the ultra-extended GRS candidate J0949+7314: our pipeline gives a measured LAS of $\sim$13.1 arcmin for this source, while its counterpart in the catalog of \cite{2024A&A...691A.185M} reports an unusually large LAS of $\sim$17.3 arcmin. This discrepancy is attributed to the fact that TGSS has significantly poorer angular resolution and sensitivity compared to LoTSS, causing some diffuse structures to be undetected in the TGSS data used by the reference catalog, thereby leading to an underestimation of the source extent. These rare outlier cases do not affect the overall robustness of our pipeline LAS output.

The bottom two panels of Figure~\ref{fig:z_las_lls_anal} present the LLS comparison results for the full sample of cross-matched GRS candidates collected from pre-existing public catalogs. The left scatter panel plots our pipeline-derived LLS ($\mathrm{LLS_{our}}$) against the literature LLS ($\mathrm{LLS_{known}}$) extracted from published GRS catalogs, with the black dashed line marking the 1:1 perfect consistency reference. Among the entire matched sample, 96.4\% of all sources have their relative LLS difference $\Delta \mathrm{LLS}/\mathrm{LLS_{our}}$ (defined as $\Delta \mathrm{LLS} = \mathrm{LLS_{known}} - \mathrm{LLS_{our}}$) distributed within the interval of $-0.5$ to $+0.5$, forming a prominent sharp central peak in the normalized difference histogram on the right panel. Only 3.6\% of sources extend into the long tails across the full axis span up to $\pm 0.75$, and 96.4\% of the data points are clustered along the 1:1 reference line, which robustly confirms that the LLS values output by our measurement pipeline show excellent overall consistency with the LLS measurements of the same matched GRSs in pre-existing catalogs. Only a small number of sources exhibit noticeable deviations far from the 1:1 reference line. As derived from the redshift consistency analysis in the previous section, these large deviations are dominantly caused by systematic differences in the adopted redshift values: even for sources with perfectly consistent measured LAS values, mismatched redshift extractions used in our pipeline and the reference catalogs will introduce non-negligible offsets in the cosmological angular diameter distance calculation, leading to distinct LLS outputs. These rare outlier cases do not affect the overall robustness of our pipeline-derived LLS results for the GRS candidate sample.

Our pipeline incorporates a duplicate removal step that retains only candidates with a confidence score exceeding 0.5. The pixel-level mask for each GRS candidate is generated using a $3\sigma_{\rm rms}$ threshold. Visual inspection confirms that the masks produced by our pipeline ensure a reasonable and effective estimation of the LAS for each GRS candidate. This conclusion is further supported by the Pearson correlation coefficient of $\rho$=0.92 between the LASs of our GRS candidates and those from existing GRS catalogs for the matched sources. Based on these results, we consider the matched GRS candidates to be highly reliable. The primary uncertainty for the unmatched GRS candidates arises from redshift extractions. Our inspection indicates that the spectroscopic redshifts are generally accurate, whereas the uncertainties are predominantly associated with the photometric redshifts. To evaluate the reliability of our catalog size constraints, we assigned three reference redshifts ($z = 0.2$, $0.3$, and $0.5$) to the unmatched GRS candidates that have photometric redshifts. Although modern machine learning--identified GRS populations typically reside at much higher redshifts \citep[e.g.,][]{2024A&A...691A.185M}, these classical values are adopted here as conservative proxies to test the lower-bound LLS and overall validity of the candidates. We then recomputed their LLS values under these three redshift assumptions. The numbers of sources that remain as GRS candidates under these fixed redshifts are 160, 610, and 1,260, respectively.

The final number of reliable GRS candidates is the sum of three components: (1) sources matched with existing GRS catalogs, (2) the remaining unmatched sources with spectroscopic redshifts, and (3) sources that satisfy the GRS criterion under the classical redshift assumptions. This yields total reliable candidate numbers of 2,187, 2,637, and 3,287, corresponding to final precision rates of 39.1\%, 47.1\%, and 58.7\%, respectively. \cite{2024A&A...691A.185M} have explicitly quantified the detection and redshift estimation efficacy of their GRS identification pipeline, reporting a precision rate of 47\%. In addition to redshift, we also compared the LAS and LLS between our pipeline and existing catalogs. Compared to the consistency rate they reported for their overlapping sources with existing catalogs, our pipeline achieves a significantly higher cross-source agreement rate. Notably, for a redshift assumption of $z=0.3$ or higher, our pipeline achieves a precision rate that exceeds the 47\% rate reported by \cite{2024A&A...691A.185M}. 
This further demonstrates the high reliability and robustness of our GRS candidate screening, redshift extraction, LAS estimation, and LLS calculation workflow.

\begin{deluxetable*}{lcccccccc}
\digitalasset
\tablewidth{0pt}
\tablecaption{Comparison of our sample with earlier GRS studies  \label{tab:GRS_comp}}
\tablehead{
\colhead{Surveys} & \colhead{Frequency} & \colhead{Number} & \colhead{${\rm LAS}_{\rm med}$} & \colhead{${F}_{\rm min,med}$} & \colhead{${\rm log}_{10}({\rm P}_{\rm min,med})$} & \colhead{$z_{\rm med}$} &\colhead{${\rm f}_{\rm zsp}$,${\rm f}_{\rm QSO}$} & \colhead{${\rm LLS}_{\rm med}$}  \\
\colhead{} & \colhead{(MHz)} & \colhead{of GRS} & \colhead{(arcmin)} & \colhead{(mJy)} & \colhead{(${\rm{W}}\,{{\rm Hz}^{ - 1}}$)} & \colhead{} & \colhead{} & \colhead{(Mpc)} 
}
\colnumbers
\startdata
FIRST,NVSS$^1$&1400 &349 &5.40 &5,164 &23.0,25.5 &0.240 &86.8\%,19.8\%&1.14  \\
FIRST,NVSS$^2$&1400 &55 &4.07 &4,40 &---,--- &0.210 &100\%,0.0\%& 0.81 \\
LoTSS$^3$&144 &239 &2.42 &2,118 &24.0,26.2 &0.534 &63.2\%,16.7\%&0.89  \\
NVSS$^4$&1400 &820 &4.33 &25,211 &23.7,25.5 &0.300 &12.3\%,2.8\%&1.05  \\
NVSS,FIRST$^5$&1400 &174 &2.10 &---,--- &24.4,26.1 &0.910 &100\%,100\%&0.92 \\
RACS$^6$&888 &181 & 3.22 &5,41 &23.5,25.9 &0.650 &7.2\%,17.7\%&1.22 \\
LoTSS$^7$&150 &74 & 2.60 &4,29 &24.5,25.8 &0.800 &31.1\%,13.5\%&1.00 \\
LoTSS$^8$&144 &2,060 & 5.15 &---,--- &---,--- &0.291 &50.2\%,2.2\%&1.35 \\
TGSS$^9$&150 &34& 2.87 &420,830 &26.0,27.1 &0.521 &73.5\%,29.4\%&1.06 \\
LoTSS$^{10}$&144 &281& 2.40 &2,28 &24.1,25.7 &0.780 &35.9\%,14.2\%&0.96 \\
LoTSS$^{11}$ &144 &11,585& 2.32 &---,--- &---,--- &0.717 &---,---&0.94 \\
Multiple$^{12}$&--- &142& 8.90 &4,86 &24.6,26.3 &0.670 &33.8\%,16.2\%&3.41 \\
TGSS$^{13}$&150 &53 & 1.92 &90,310 &25.9,26.7 &0.977 &100\%,100\%&0.84 \\
TGSS (this work)&150 &5,595 & 2.00 &20,500 &25.2,27.1 &0.875 &25.6\%,18.4\%&0.90 \\
\enddata
\tablecomments{Column (1): surveys used; flags represent the following references: 1=\cite{2018ApJS..238....9K}, 2=\cite{2020ApJS..247...53K}, 3=\cite{2020A\string&A...635A...5D}, 4=\cite{2020A\string&A...642A.153D}, 5=\cite{2021ApJS..253...25K}, 6=\cite{2021Galax...9...99A}, 7=\cite{2022MNRAS.515.2032S}, 8=\cite{2023A\string&A...672A.163O}, 9=\cite{2024ApJS..273...30B}, 10=\cite{2024A\string&A...686A..21S}, 11=\cite{2024A\string&A...691A.185M}, 12=\cite{2025A\string&A...699A.257A}, and 13=\cite{2025ApJS..281...34M}. ``Multiple" indicates that the GRS samples come from various modern surveys: NVSS, FIRST, LoTSS, TGSS, RACS, the GaLactic and Extragalactic All-sky Murchison Widefield Array (GLEAM; \cite{2017MNRAS.464.1146H}), the GaLactic and Extragalactic All-sky Murchison Widefield Array eXtended (GLEAM-X; \cite{2022PASA...39...35H}), the Evolutionary Map of the Universe (EMU; \cite{2021PASA...38...46N}), the MeerKAT Absorption Line Survey (MALS; \cite{2024ApJS..270...33D}), and so on. Column (2): observed frequency in surveys. Column (3): the number of GRS published. Column (4): median of LAS values. Column (5): minimum and median integrated flux density of the GRS. Column (6): minimum and median decadic logarithm of the radio power at column (2) frequency. Column (7): median redshift of the GRS hosts. Column (8): hosts fractions with spectroscopic redshift and of those that are QSOs or candidates. Column (9): median of LLS values.}
\end{deluxetable*}

\section{Discussion} \label{Sec:Disc}


\subsection{A Comparison Between Our Sample and Earlier GRSs}
Table~\ref{tab:GRS_comp} compares our sample with several previously published GRS catalogs. Prior to this work, the largest individual samples were those from \cite{2024A&A...691A.185M} and \cite{2023A\string&A...672A.163O}, containing 11,585 and 2,060 GRSs, respectively. Other notable contributions include NVSS-based GRSs from \cite{2020A\string&A...642A.153D} (820) and FIRST-based GRSs from \cite{2018ApJS..238....9K} (349). By combining these existing efforts, the total number of previously known GRSs across these major surveys was significant, yet often fragmented across different frequencies and sky coverages. Our new catalog of 5,595 GRS candidates from the TGSS includes 4,566 previously unknown sources, representing a substantial increase in the total known population. This complements the recent catalog of \cite{2024A&A...691A.185M} (11,585 GRSs), and together these works have significantly advanced the census of GRSs.

The comparative data in Table~\ref{tab:GRS_comp} reveal several key trends. While early catalogs like \cite{2025A&A...699A.257A} and \cite{2023A&A...672A.163O} targeted sources with larger median LLS (${\rm LLS}_{\rm med}$ of 3.41 Mpc and 1.35 Mpc, respectively), our sample presents a more moderate value of $\sim$0.9~Mpc. This places our sample in a similar physical scale regime to \cite{2021ApJS..253...25K} (0.92 Mpc) and \cite{2020A&A...635A...5D} (0.89 Mpc). In terms of cosmic distance, our sample reaches a median redshift ($z_{\rm med}$) of 0.875, which is considerably higher than most previous low-frequency surveys, surpassed only by the specialized \cite{2025ApJS..281...34M} and \cite{2021ApJS..253...25K} samples. Notably, our catalog achieves such a high redshift while maintaining a lower flux density (${F}_{\rm min}$=20 mJy), demonstrating a superior ability of our pipeline to detect faint, high-redshift GRSs compared to earlier TGSS-based efforts like \cite{2024ApJS..273...30B}.

\begin{deluxetable*}{lccccccccc}
\digitalasset
\tablewidth{0pt}
\tablecaption{Galaxy cluster properties for 286 GRS candidates  \label{tab:cluster}}
\tablehead{
\colhead{Short Name} & \colhead{Cluster} & \colhead{z$_{Cl}$} & \colhead{$D_c$} & \colhead{Sep} & \colhead{$D_{l}$} & \colhead{$r_{500}$} & \colhead{$R_{L*,500}$} & \colhead{$M_{500}$} & \colhead{$N_{500}$} \\
\colhead{} &\colhead{} & \colhead{}  & \colhead{(Mpc)} & \colhead{(arcsec)} & \colhead{(kpc)} & \colhead{(Mpc)} & \colhead{} & \colhead{($\times10^{14}M_{\odot}$)} & \colhead{}  
}
\colnumbers
\startdata
J0020+1816&CFSFDP J002057.5+181611 &0.5163$^{\rm S}$ &1942 &0 &0 &0.53 &17.40 &0.8 &12 \\
J0022-0818&  Y21 J002225.0-081846&  0.5714$^{\rm S}$&  2117&  0&  0&  0.61&  27.39&  1.2&  9 \\
J0023-1422&  WH J002314.9 142240&  0.4868$^{\rm P}$&  1800&  31&  193&  0.40&  11.29&  0.5&  7 \\
J0023-2502&  CFSFDP J002308.9-250224&  0.3504$^{\rm P}$&  1400&  6&  31&  0.50&  16.96&  0.8&  9 \\
J0026-1218&  Y21 J002659.9-121838&  0.1598$^{\rm P}$&  660&  0&  0&  0.78&  27.90&  1.3&  17 \\
J0027+1632&  CFSFDP J002712.8+163227&  0.3785$^{\rm S}$&  1476&  0&  0&  0.64&  27.60&  1.2&  9 \\
J0028+0035&  WHL J002838.9+003540&  0.3985$^{\rm S}$&  1546&  15&  83&  0.63&  19.16&  0.9&  9 \\
J0029+1842&  WH J002915.4+184343&  0.3882$^{\rm S}$&  1510&  49&  267&  0.51&  13.01&  0.6&  6 \\
J0033+0243&  AMF J003339.2+024321&  0.4697$^{\rm S}$&  1789&  0&  0&  1.04&  116.18&  5.0&  83 \\
J0034+0904&  WH J003448.3+090419&  0.6948$^{\rm S}$&  2491&  0&  0&  0.76&  55.18&  2.5&  8 \\
\multicolumn{1}{c}{...} & ... & ... & ...  & ... & ... & ... & ... & ... & ... \\
\enddata
\tablecomments{Column (1): short name (JHHMM+DDMM) of the GRS candidates. Column (2): cluster name with J2000 coordinates. Column (3): cluster redshift, with flags: `P' represents photometric redshift, `S' represents spectroscopic redshift. Column (4): comoving distance (in Mpc). Column (5): angular separation between GRS candidates and their associated cluster centres (in arcsec). Column (6): sky-projected distances between the positions of GRS hosts and the centres of their associated clusters (in kpc). Column (7): cluster radius (in Mpc). Column (8): cluster richness. Column (9): cluster mass ($\times10^{14}M_{\odot}$). Column (10): number of member galaxy candidates within $r_{500}$. Table \ref{tab:cluster} is published in its entirety in the electronic edition of \href{https://doi.org/10.3847/1538-4365/ae96ae}{\textit{the Astrophysical Journal Supplements}}. A portion is shown here for guidance regarding its form and content.}
\end{deluxetable*}

\subsection{Association of Our Sources with Galaxy Clusters}
The relationship between GRSs and their large-scale environments remains a subject of active debate, specifically whether these giants preferentially reside in filaments, galaxy groups, or galaxy clusters \citep{2024A&A...686A.137O}. In this section, we examined whether the GRS candidates in our catalog were possibly associated with known galaxy clusters from DESI LS DR10 \citep{2024ApJS..272...39W} and SDSS \citep{2015ApJ...807..178W}. It is important to note that both catalogs include a significant number of galaxy clusters known prior to their construction. We used these two galaxy cluster catalogs because they almost covered the entire sky region of our GRS candidate catalog. Our association procedure first cross-matched our GRS catalog with each of the two galaxy cluster catalogs using a radius of 3$'$. For each matched source, we then determined whether the GRS host is a genuine cluster member by checking two criteria: (1) whether the separation is less than or equal to the cluster radius ($r_{500}$), and (2) whether the absolute difference between the redshift of the GRS host and that of the galaxy cluster ($z_{Cl}$) is less than 0.01. Following this procedure, we identified a total of 286 GRS candidates as cluster members, representing 5.1\% of the total GRS candidate sample. Among these, 275 were obtained from \cite{2024ApJS..272...39W} and 11 from \cite{2015ApJ...807..178W}. This fraction is significantly lower than the $\sim$23\% found in the local Universe ($z<0.16$) by \cite{2024A&A...686A.137O}. However, the median redshift of our GRS candidates is $z=0.875$, substantially higher than that of \cite{2024A&A...686A.137O}. This tension probably arises from observational selection effects: cluster catalogues become increasingly incomplete at higher redshifts (especially for group-mass halos), and massive structures were less abundant at earlier cosmic epochs. Consequently, our derived fraction of 5.1\% should be regarded as a conservative lower limit.
Detailed properties of the associated galaxy clusters for the 286 GRS candidates are listed in Table~\ref{tab:cluster}. These clusters originate from various catalogs: 93 from \cite{2024ApJS..272...39W}, 55 from \cite{2015ApJ...807..178W}, 63 from \cite{2021ApJ...909..143Y}, 7 from \cite{2018MNRAS.475..343W}, 3 from \cite{2022MNRAS.513.3946W}, 4 from \cite{2021MNRAS.500.1003W}, 6 from \cite{2007ApJ...660..239K}, 5 from \cite{2010ApJS..191..254H}, 2 from \cite{2024MNRAS.531.2285Y}, 27 from \cite{2021ApJS..253...56Z,2022RAA....22f5001Z}, 4 from \cite{2014MNRAS.444..147O}, and 17 from \cite{2011ApJ...736...21S}. Among these 286 GRS candidates, 93.7\% (268/286) are located at $z<$0.9, indicating that this subsample is primarily composed of sources in the relatively late Universe, where cluster memberships are well-characterized. Furthermore, we found that the LLSs of these 286 GRS candidates are all less than 3 Mpc, which is consistent with the view of \citet{2024A&A...687L...8S} that GRSs with LLS exceeding $\sim$3 Mpc are not typically associated with galaxy clusters. The host galaxy clusters of these 286 GRS candidates span a wide mass range, from 0.5$\times10^{14}M_{\odot}$ to 6.1$\times10^{14}M_{\odot}$ with a mean mass of 1.2$\times10^{14}M_{\odot}$. This distribution indicates that GRSs can reside in diverse environments, ranging from galaxy groups to massive clusters. The presence of GRSs in such varied gravitational potentials suggests that their giant scales are not solely a result of low-density environments, but are likely driven by the intrinsic properties of the central engine.

\section{Conclusions} \label{Sec:Conc}
The physical mechanisms enabling GRSs to reach large scales, as well as the role of their surrounding cosmic environments, remain central topics of investigation in extragalactic astrophysics. Traditionally, the identification of GRSs has relied heavily on visual inspection, which, despite its reliability, is remarkably inefficient when applied to modern large-scale radio surveys. With the advent of the next generation of sensitive, wide-area surveys, the manual construction of large GRS catalogs has become a significant bottleneck. To address the limitations of manual identification, we developed a deep learning--based pipeline for the systematic and automated search of GRSs within the TGSS ADR1 dataset. By applying this pipeline, we have constructed a catalog of 5,595 GRS candidates, consisting of 4,210 GRGs and 1,030 GRQs. The remaining 355 sources remain unclassified due to the lack of definitive classification information for their host galaxies in existing spectroscopic catalogs. Notably, 4,566 GRS candidates in our sample are newly identified, marking a substantial increase in the total number of known GRSs to date.

The constructed GRS candidate catalog covers a redshift range from $z = 0.056$ to at least $z = 2.385$, with a median of $z = 0.875$. The spectral indices of our GRS candidates ($\alpha^{\rm 1400}_{\rm 150}$) with a median of $\alpha^{\rm 1400}_{\rm 150}$=-0.75, are slightly steeper than the typical value of -0.7 for extended radio sources. Similar to previous TGSS-based GRS samples, our GRS candidates exhibit high radio powers at 150 MHz, ${\rm log}_{10}(P_{150}/\rm W\,Hz^{-1})$, ranging from $\sim$25.2 to $\sim$30.0 with a mean and median of 27.2 and 
27.1, respectively. The LLS of the sample range from $\sim$0.7 Mpc to $\sim$4.5 Mpc. Most GRS candidates show slightly bent morphologies, with a mean and median BA of 14.3$^\circ$ and 11.0$^\circ$. Their $M_r$ 99.2\% lie within $-27 \lesssim M_r \lesssim -17$ mag. By cross-matching with existing galaxy cluster catalogs, we identify 286 GRS candidates as galaxy cluster members. This finding suggests that the formation of GRSs is not solely due to low-density environments but may also be related to unique central engine activities. The catalog presented here provides a valuable sample for investigating the physical properties and environments of GRSs. Furthermore, the deep learning--based pipeline developed in this work offers a promising tool for searching for GRSs in future large-scale radio surveys, such as those conducted by the SKA \citep{2009IEEEP..97.1482D} and the Deep Synoptic Array \citep[DSA; e.g.][]{2021AAS...23731604R} at 1400~MHz.

\begin{acknowledgments}
This work was supported by the National SKA Program of China (Nos. 2025SKA0160100, 2022SKA0120101, 2022SKA0130100, 2022SKA0130104), the National Natural Science Foundation of China (Nos. 12573101, 12103013), the Youth Innovation Promotion Association, Chinese Academy of Sciences(Grant No. 2021258), the INTERNATIONAL PARTNERSHIP PROGRAM OF THE CHINESE ACADEMY OF SCIENCES (No. 018GJHZ2024025GC), the Foundation of Science and Technology of Guizhou Province (Nos. (2021)023), the Foundation of Guizhou Provincial Education Department (Nos. KY(2021)303, KY(2020)003, KY(2023)059), and the Guangxi Universities Engineering Research Center for Optoelectronic Information Technology. This work used resources of China SKA Regional Centre funded by Ministry of Science and Technology of the People’s Republic of China. This work is supported by 100101 State Key Laboratory of Radio Astronomy and Technology (Chinese Academy of Sciences).

We thank the staff of the GMRT that made these observations possible. GMRT is run by the National Centre for Radio Astrophysics of the Tata Institute of Fundamental Research.

This research used data obtained with the Dark Energy Spectroscopic Instrument (DESI). DESI construction and operations is managed by the Lawrence Berkeley National Laboratory. This material is based upon work supported by the U.S. Department of Energy, Office of Science, Office of High-Energy Physics, under Contract No. DE-AC02-05CH11231, and by the National Energy Research Scientific Computing Center, a DOE Office of Science User Facility under the same contract. Additional support for DESI was provided by the U.S. National Science Foundation (NSF), Division of Astronomical Sciences under Contract No. AST-0950945 to the NSF’s National Optical-Infrared Astronomy Research Laboratory; the Science and Technology Facilities Council of the United Kingdom; the Gordon and Betty Moore Foundation; the Heising-Simons Foundation; the French Alternative Energies and Atomic Energy Commission (CEA); the National Council of Humanities, Science and Technology of Mexico (CONAHCYT); the Ministry of Science and Innovation of Spain (MICINN), and by the DESI Member Institutions: www.desi.lbl.gov/collaborating-institutions. The DESI collaboration is honored to be permitted to conduct scientific research on I’oligam Du’ag (Kitt Peak), a mountain with particular significance to the Tohono O’odham Nation. Any opinions, findings, and conclusions or recommendations expressed in this material are those of the author(s) and do not necessarily reflect the views of the U.S. National Science Foundation, the U.S. Department of Energy, or any of the listed funding agencies.

The DESI Legacy Imaging Surveys consist of three individual and complementary projects: the Dark Energy Camera Legacy Survey (DECaLS), the Beijing-Arizona Sky Survey (BASS), and the Mayall z-band Legacy Survey (MzLS). DECaLS, BASS and MzLS together include data obtained, respectively, at the Blanco telescope, Cerro Tololo Inter-American Observatory, NSF’s NOIRLab; the Bok telescope, Steward Observatory, University of Arizona; and the Mayall telescope, Kitt Peak National Observatory, NOIRLab. NOIRLab is operated by the Association of Universities for Research in Astronomy (AURA) under a cooperative agreement with the National Science Foundation. Pipeline processing and analyses of the data were supported by NOIRLab and the Lawrence Berkeley National Laboratory (LBNL). Legacy Surveys also uses data products from the Near-Earth Object Wide-field Infrared Survey Explorer (NEOWISE), a project of the Jet Propulsion Laboratory/California Institute of Technology, funded by the National Aeronautics and Space Administration. Legacy Surveys was supported by: the Director, Office of Science, Office of High Energy Physics of the U.S. Department of Energy; the National Energy Research Scientific Computing Center, a DOE Office of Science User Facility; the U.S. National Science Foundation, Division of Astronomical Sciences; the National Astronomical Observatories of China, the Chinese Academy of Sciences and the Chinese National Natural Science Foundation. LBNL is managed by the Regents of the University of California under contract to the U.S. Department of Energy. The complete acknowledgments can be found at https://www.legacysurvey.org/acknowledgment/.

Funding for the Sloan Digital Sky 
Survey IV has been provided by the 
Alfred P. Sloan Foundation, the U.S. 
Department of Energy Office of 
Science, and the Participating 
Institutions. 

SDSS-IV acknowledges support and 
resources from the Center for High 
Performance Computing  at the 
University of Utah. The SDSS 
website is www.sdss4.org.

SDSS-IV is managed by the 
Astrophysical Research Consortium 
for the Participating Institutions 
of the SDSS Collaboration including 
the Brazilian Participation Group, 
the Carnegie Institution for Science, 
Carnegie Mellon University, Center for 
Astrophysics | Harvard \& 
Smithsonian, the Chilean Participation 
Group, the French Participation Group, 
Instituto de Astrof\'isica de 
Canarias, The Johns Hopkins 
University, Kavli Institute for the 
Physics and Mathematics of the 
Universe (IPMU) / University of 
Tokyo, the Korean Participation Group, 
Lawrence Berkeley National Laboratory, 
Leibniz Institut f\"ur Astrophysik 
Potsdam (AIP),  Max-Planck-Institut 
f\"ur Astronomie (MPIA Heidelberg), 
Max-Planck-Institut f\"ur 
Astrophysik (MPA Garching), 
Max-Planck-Institut f\"ur 
Extraterrestrische Physik (MPE), 
National Astronomical Observatories of 
China, New Mexico State University, 
New York University, University of 
Notre Dame, Observat\'ario 
Nacional / MCTI, The Ohio State 
University, Pennsylvania State 
University, Shanghai 
Astronomical Observatory, United 
Kingdom Participation Group, 
Universidad Nacional Aut\'onoma 
de M\'exico, University of Arizona, 
University of Colorado Boulder, 
University of Oxford, University of 
Portsmouth, University of Utah, 
University of Virginia, University 
of Washington, University of 
Wisconsin, Vanderbilt University, 
and Yale University.

This research uses services or data provided by the Astro Data Lab, which is part of the Community Science and Data Center (CSDC) Program of NSF NOIRLab. NOIRLab is operated by the Association of Universities for Research in Astronomy (AURA), Inc. under a cooperative agreement with the U.S. National Science Foundation.

This research work made use of public data from the Karl G. Jansky Very Large Array (VLA); the VLA facility is operated by the National Radio Astronomy Observatory (NRAO). The NRAO is a facility of the National Science Foundation operated under cooperative agreement by Associated Universities, Inc. 

This research has made use of the NASA/IPAC Extragalactic Database (NED), which is funded by the National Aeronautics and Space Administration and operated by the California Institute of Technology.

\end{acknowledgments}





%
\facilities{GMRT, CDS, VLA, WISE, Sloan, Astro Data Lab}

\software{astropy \citep{2013A&A...558A..33A,2018AJ....156..123A,2022ApJ...935..167A}, APLpy \citep{2012ascl.soft08017R}, Matplotlib \citep{2007CSE.....9...90H}, Astroquery \citep{2019AJ....157...98G}, Numpy \citep{2020Natur.585..357H}, Pandas \citep{mckinney-proc-scipy-2010,reback2020pandas}, scikit-image \citep{van2014scikit}, SciPy \citep{virtanen2020scipy}, SAOImage DS9 \citep{2003ASPC..295..489J}, TOPCAT \citep{2005ASPC..347...29T}}

\bibliography{sample701}{}
\bibliographystyle{aasjournalv7}



\end{document}